\documentclass[fleqn,usenatbib]{mnras}

\usepackage{newtxtext,newtxmath}
\usepackage{xcolor}
\usepackage[dvipsnames]{xcolor}
\usepackage{booktabs, caption} 

\usepackage[T1]{fontenc}
\usepackage{float} 
\DeclareRobustCommand{\VAN}[3]{#2}
\let\VANthebibliography\thebibliography
\def\thebibliography{\DeclareRobustCommand{\VAN}[3]{##3}\VANthebibliography}

\usepackage{enumerate}

\usepackage{graphicx, float}	
\usepackage{amsmath}	
\renewcommand{\vec}[1]{\boldsymbol{#1}} 

\title[Eddington-limited Emission from TDEs]{Wind-mediated Eddington-limited emission in a $10^4 M_\odot$ Black Hole Tidal Disruption Event}

\author[P. Martire et al.]{
P. Martire$^{1}$\thanks{E-mail: martire@strw.leidenuniv.nl}, 
E.M. Rossi,$^{1}$
N. C. Stone,$^{2,3}$
E. Steinberg,$^{3}$,
K. Kilmetis$^{1}$,
I. Linial$^{4}$
\\
$^{1}$Leiden Observatory, Leiden University, PO Box 9513, 2300 RA Leiden, The Netherlands \\
$^{2}$Department of Astronomy, University of Wisconsin, Madison, WI 53706, USA\\
$^{3}$Racah Institute of Physics, The Hebrew University, Jerusalem, 91904, Israel\\
$^{4}$Department of Physics and Columbia Astrophysics Laboratory, Columbia University, New York, NY 10027, USA\\
}
\date{Accepted XXX. Received YYY; in original form ZZZ}

\pubyear{\the\year{}}

\begin{document}
\label{firstpage}
\pagerange{\pageref{firstpage}--\pageref{lastpage}}
\maketitle

\begin{abstract}
Observations of tidal disruption events (TDEs) have already produced tens of strong candidate flares, and their number will greatly increase with upcoming wide field surveys.  Nevertheless, the origin of the measured luminosity peak at early times is still unknown, and the ultimate sources of energy dissipation in TDEs are not fully understood. 
Here we present the first three-dimensional end-to-end simulation of a TDE by a $10^4 M_\odot$ intermediate mass black hole (IMBH) with realistic parameters, run with the radiation-hydrodynamics code \texttt{RICH}. 
We find that the stellar debris fails to circularize efficiently, while a low-density, radiation-driven wind forms near pericenter and expands quasi-spherically. 
Radiation is advected by this outflow and released at the photosphere, which expands to radii of $\approx10^{13}$ cm and reaches temperatures of $\sim$few $\times10^4$K at the peak of the light curve. The resulting luminosity briefly exceeds the Eddington limit before settling near that value. We systematically test the numerical convergence of our simulation by running it at three resolutions. While the nozzle shock at pericenter may be under-resolved, we find that global results are qualitatively converged and, largely, quantitatively robust. The upcoming Vera Rubin Observatory's LSST (g and r band) and {\it ULTRASAT} (near UV) will be able to observe events like our simulated IMBH TDE up to redshifts of $z\approx0.1$ and $z\approx0.06$, respectively.
\end{abstract}




\section{Introduction}
Tidal disruption events (TDEs) are among the brightest thermal transients in the optical \citep{Sjoert11, Gezari12, Sjoert21}, UV \citep{Gezari08, SjoertUV}, and X-ray \citep{Komossa99, Komossa15, Sazonov21, Saxton21, Grotova25} bands. 
They occur when a star comes close enough to a Massive Black Hole (MBH) to be torn apart by tidal forces \citep{Hills75, LuminetCarter, 
Rees88} and they release energy with luminosities comparable to those of the brightest supernovae \citep{Arcavi14, Leloudas16}.
TDEs can be used as probes of {\emph quiescent} MBHs in nearby galaxies, and are especially interesting for Intermediate-Mass Black Holes (IMBHs), i.e. BHs with mass $M_\text{BH}\in[10^4,10^6]M_\odot$, for which other observational techniques, such as dynamical mass measurements, are limited to the nearby Universe and to a handful of targets \citep{Greene20}. Currently, demographics and even the existence of IMBHs remain a topic of debate  \citep{Greene20, Patra25}.

By detecting and studying TDEs, we can confirm the presence of MBHs and potentially estimate their masses \citep{Guillochon14, Mockler19, Ryu20} and spins \citep{Stone12, Gafton19, Wen20, Mummery24}. Mass estimates would allow us to probe the relation $M-\sigma$ \citep{Msigma} in the lower-mass regime and would provide information on the formation of  MBHs \citep{Volonteri10, Bhowmick24}; likewise, spin measurements would inform questions regarding the growth histories of MBHs \citep{Berti08}. 

TDEs are also interesting as probes of accretion physics, such as thermal/viscous stability \citep{Shen14, Kaur23, Piro25}, time-dependent accretion flows \citep{Cannizzo90, vanVelzen19, Mummery19, Tamilan24, Alush25}, the physics of inclined \citep{Stone12, Franchini16, Zanazzi19} or eccentric \citep{Zanazzi20} accretion discs, and jet launching physics \citep{Stone12, Tchekhovskoy14, Liska18, Coughlin20}. TDEs involving IMBHs are especially noteworthy because they can achieve uncommonly large fallback rates (e.g., $\sim 10^5$ the Eddington accretion rate for $M_{\rm BH}=10^4 M_\odot$; \citealt{stone13}), logarithmically intermediate between gamma-ray bursts ($\approx 10^{10-14}$ the photon Eddington rate) and active galactic nuclei and X-ray bursts ($\lesssim$few times the Eddington rate).  Other known super-Eddington sources generally have accretion rates orders of magnitude below the theoretical IMBH TDE fallback rate (e.g., ultraluminous X-ray sources, thought to accrete at $\lesssim 100$ times the Eddington rate; \citealt{Socrates06}).  

Candidate IMBHs-TDEs have been claimed, especially through X-ray detection \citep{Maksym14, Lin18, Wen21, Sazonov21, Cao23, Jin25}. Additionally, other observed transient classes of unknown origin, such as Luminous Fast Blue Optical Transients (LFBOTs) and Fast X-ray Transients (FXTs), have been proposed to be IMBH-TDEs \citep{Perley19, Angus22, Migliori24, Cao24, Inkenhaag25}. Despite the increase of  observations, the lack of a robust model describing the early-time thermal emission is preventing us to realise their potential. 

Although the star's disruption is reasonably well understood \citep{Rossi21}, subsequent debris evolution is still incompletely explained. In the classical picture, half of the stellar mass is thrown to infinity after the disruption, while the other half remains gravitationally bound. The latter returns to the vicinity of the MBH and starts to circularise around it, emitting light from the recently formed compact accretion disc. 
Although X-ray spectra are often consistent with this picture, UV and optical observations \citep{Sjoert11, Gezari12, Arcavi14} infer lower blackbody temperatures and bolometric luminosities and larger blackbody radii ($T_\text{BB}\approx 2-5\times 10^4 ~{\rm K}, \,L\sim10^{43-45} \text{erg/s}, \,R_\text{BB}\sim 10^{15}\text{cm}$) than theoretical predictions for compact, fully circularised discs \citep{Ulmer99, Strubbe09}. Two general mechanisms have been suggested to understand this mismatch and to explain optical/UV emission in TDEs, especially at early times. 
A reprocessing layer converting the X-ray radiation from the inner disc into longer wavelength photons could explain optical/UV observations \citep{Loeb97,CoughlinBeg14, Guillochon14, Metzger16, Roth16}, producing Eddington-limited luminosities. An alternative explanation involves shocks, which could explain the observed optical/UV luminosities even without accretion, due to dissipation in and between debris streams \citep{Piran15, Ryu23, SS24}.  
Unlike the reprocessed accretion power model, shock-driven emission is not necessarily subject to the Eddington limit \citep{Shio15, BonnerotLu21, Huang24}.

Numerical simulations have become fundamental for investigating TDE debris evolution, serving as a bridge between analytical models and observations. Despite recent advances, computational challenges related to the resolution of critical hydrodynamic processes remain. One major difficulty is the great dynamic range involved, which spans the gravitational radius $r_\text{g} \equiv GM_\text{BH}/c^2$, up to the apocenter distances $\gtrsim 10^4 r_\text{g}$. Simulations must achieve high spatial and temporal resolution to capture both the global orbital evolution and the formation of thin debris streams, accurately treating key physical ingredients such as general relativity, stellar self-gravity, the internal structure of the disrupted star, shocks, and radiative transfer. 
While some earlier works reduced the dynamical range and computational time by adopting unphysical parameters (e.g., bound highly eccentric orbits in \citet{Hayasaki13, Bonnerot16, Hayasaki16, Sad16, Liptai19, Hu24} or small MBH-star mass ratio in \citet{Ramirez-Ruiz09, Guillochon14, Shio15}), recent advances have overcome computational barriers, leading to the first \emph{end-to-end} (i.e., from the disruption to subsequent evolution over more than one fallback time) three-dimensional (3D) simulations \citep{Ryu23, Price24, SS24}.

Although the current generation of global simulations has many results in common \citep{Krolik25}, significant limitations still exist when attempting to resolve the physical scales at pericenter, where intense vertical compression of the returning stream produces the \emph{nozzle shock} \citep{Guillochon14}. In Smoothed Particle Hydrodynamics simulations, the resolution is limited because, at any instant, the number of particles passing through the nozzle is at most of order unity unless the total particle number is extremely large\footnote{This affects in particular the 
computation of hydrodynamic accelerations, which require a population of nearest neighbour particles.}. Grid-based methods can mitigate this issue by enforcing a more refined grid near pericenter, though this comes at increased computational cost. 
Eulerian grid codes are often challenged by advection errors, as returning debris streams are highly supersonic, with Mach numbers $\sim 10^2$. 
Another possible source of artifacts in 3D simulations is artificial dissipation of orbital energy. To avoid this, high-resolution two-dimensional studies have been used to study pericentric compression  \citep{BonnerotLu22}, although neglecting the full 3D geometry and the interaction between successive stream elements.\\

In this paper, we present the first end-to-end 3D radiation-hydrodynamics simulation of a main sequence star disrupted by an IMBH. Choosing an IMBH allows us to explore this enigmatic class of astrophysical transients. In addition, it offers greater computational tractability, which makes it possible to conduct more rigorous convergence tests by comparing runs with different resolutions.
The structure of the paper is as follows: in Sec.\ref{sec: meth} we present our radiation-hydrodynamical simulations. We show the results in Sec.\ref{sec: res} and discuss them in Sec.\ref{sec: disc}. In Sec.\ref{sec:concl} we present the limitations of this work and draw our main conclusions.

\section{Methods}
\label{sec: meth}

We simulated the disruption of a star (with radius $r_\star = 0.47R_\odot$, mass $M_\star=0.5M_\odot$ and polytropic index $n=1.5$) by a non-spinning BH with mass $M_\text{BH}=10^4M_\odot$ in 3D radiation-hydrodynamics. The IMBH is positioned at the origin of the reference frame, while the star is initially placed at $3r_\text{t}$ on a parabolic orbit, where $r_{\rm t}= r_\star(M_\text{BH}/M_\star)^{1/3}\approx13R_\odot$ is the tidal radius. The pericenter $r_\text{p}$ of the star is placed at $(x,y,z)=(r_\text{t}, 0,0)$, and thus the penetration factor is $\beta\equiv r_\text{t}/r_\text{p}=1$. 
We used the radiation-hydrodynamic code \texttt{RICH} \citep{rich} with a Paczyński–Wiita potential \citep{pseudopot} in order to approximate the leading-order effect of relativistic apsidal precession. 
For reasons of computational efficiency, the gravitational force is artificially smoothed within $r_0 = 0.6 r_\text{p}$ as 
\begin{equation}
\label{eq:F PW}
    \textbf{F} = -\frac{GM_\text{BH}}{r_0(r_0-2r_\text{g})^2} \textbf{r},
\end{equation}
with $\textbf{r}$ being the spherical radial vector. The validity of this choice is discussed in Sec.\ref{sec: grav}. 
Radiation is treated according to a flux limited diffusion (FLD) scheme under the approximation of grey transport; gas thermodynamics evolve via realistic equations of state that include ionisation and recombination of H and He \citep{Tomida15}. 

\begin{figure}
    \centering
    \begin{minipage}{\linewidth}
    \centering
    \includegraphics[width=\linewidth]{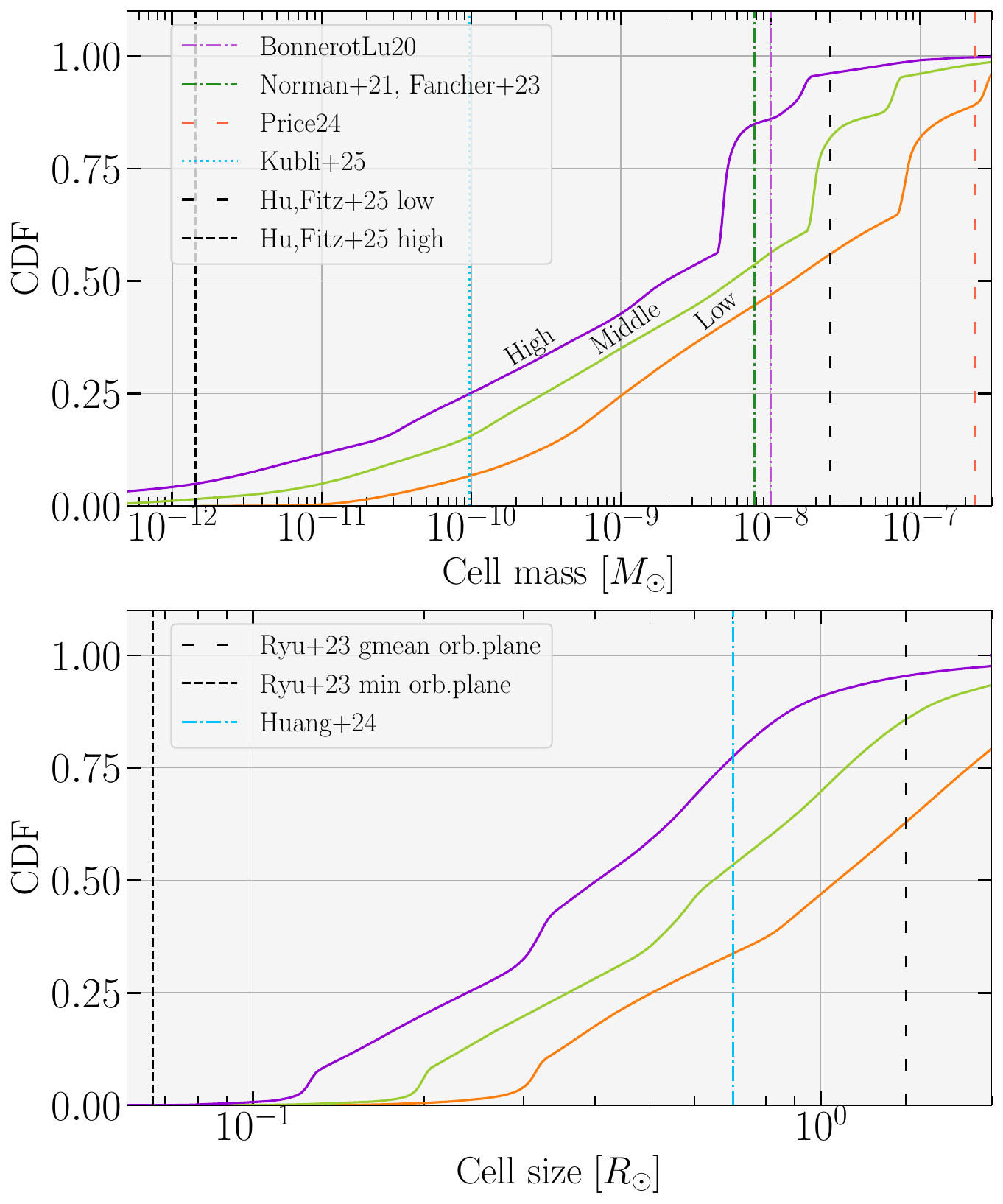}
    \caption{Cumulative distribution function (CDF) at the last common time available among resolutions ($t = 1.6 t_\text{fb}$). Orange, green and violet solid lines are respectively for \emph{Low}, \emph{Middle} and \emph{High} resolution. \emph{Upper panel}: CDF of cell mass. Vertical lines show the resolution of previous 3D simulations by \citet{BonnerotLu20, Norman21, Fancher23, Price24, Kubli25} and \citet{Hu25} (with long and short dashed vertical lines representing, respectively, its lowest and highest resolution). \emph{Lower panel}: CDF of cell size.  Vertical lines show the resolution of previous 3D simulations by \citet{Ryu23} (with long and short dashed vertical lines representing, respectively, the geometric mean and the minimum of cells size in the orbital plane) and \citet{Huang24}. We achieved unprecedented average resolution for end-to-end TDEs simulations with respect to previous works.}
    \label{fig:hist}
    \end{minipage}
\vspace{1em}
    \begin{minipage}{\linewidth}
    \centering
    \begin{tabular}{llll}
        & \textbf{Low} &\textbf{Middle} &\textbf{High}\\
        \hline\\
        min cell mass $[M_\odot]$ &$1.6\cdot10^{-12}$ &$3.6\cdot10^{-14}$ &$1.5\cdot10^{-15}$\\
        $p_{50}$ cell mass $[M_\odot]$ &$1.3\cdot10^{-8}$ &$5.5\cdot10^{-9}$ &$2.0\cdot10^{-9}$\\
        $p_{90}$ cell mass $[M_\odot]$ &$2.5\cdot10^{-7}$ &$6.4\cdot10^{-8}$ &$1.5\cdot10^{-8}$\\
        min cell size $\,\,\,[R_\odot]$ &0.120 &0.076&0.041\\
        $p_{50}$ cell size $\,\,\,[R_\odot]$ &1.1&0.65&0.40\\
        $p_{90}$ cell size $\,\,\,[R_\odot]$  &2.8 &1.6 &0.96\\ 
    \end{tabular}
    \captionof{table}{Features of the numerical resolution in our three runs, \emph{Low, Middle, and High} (shown as columns). From first to last row: minimum, $50^\text{th}$ and $90^\text{th}$ percentile of cell mass, minimum, $50^\text{th}$ and $90^\text{th}$ percentile of cell size at the last common time available among resolutions ($t = 1.6 t_\text{fb}$).}\label{tab: res}
    \end{minipage}
\end{figure}
\begin{figure*}
    \centering
    \includegraphics[width=\linewidth]{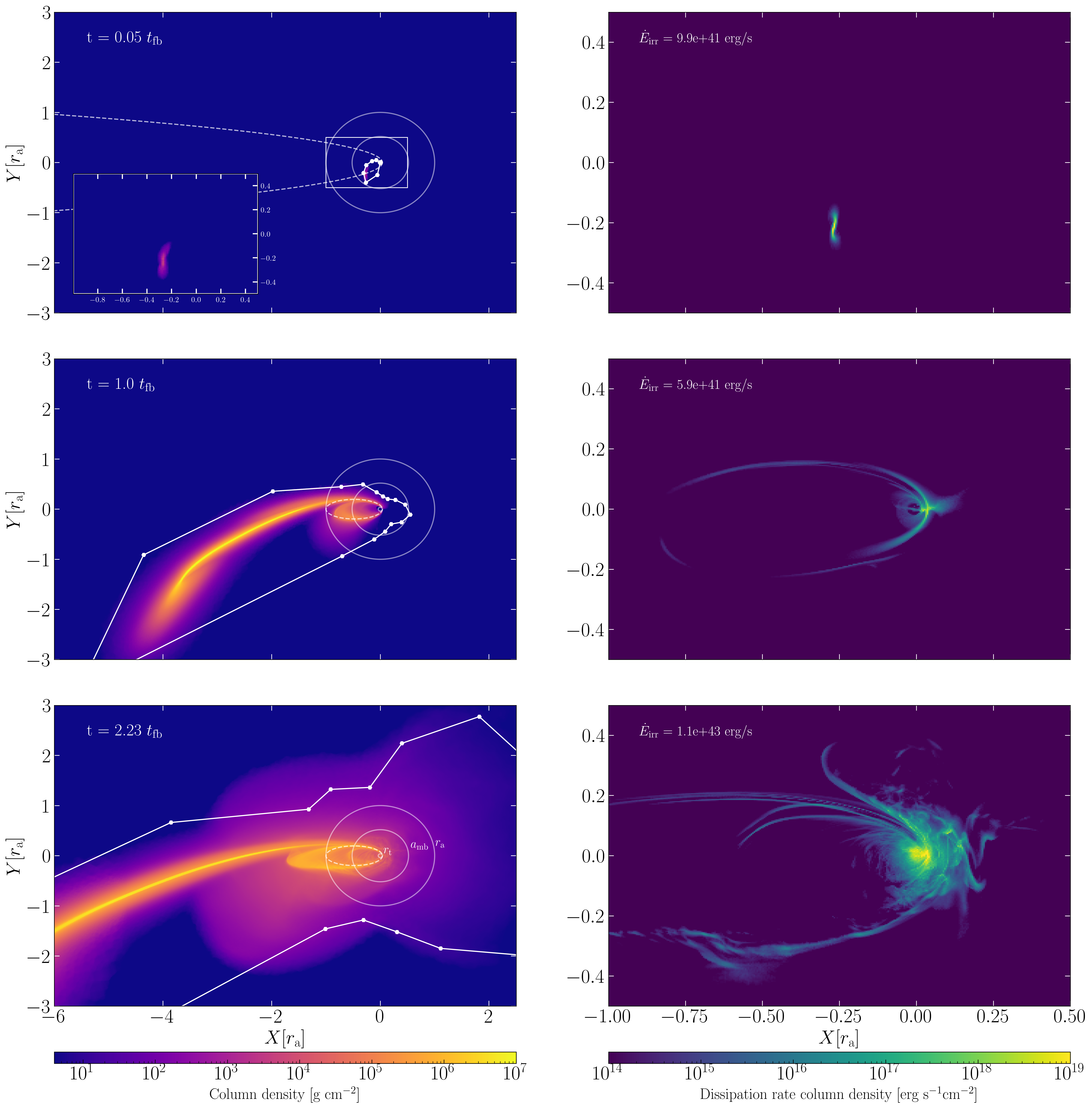}
    \caption{Mass column density (\emph{left panel}) and dissipation rate column density (\emph{right panel}) in the \emph{High res} at the onset of mass return (\emph{top panel}, with a zoom-in insert), after one fallback time $t_{\rm fb}$ (\emph{middle panel}) and at the last time available $t=2.2t_{\rm fb}$ (\emph{bottom panel}). The volume integrated dissipation rates $\dot{E}_{\rm irr}$ are reported in the right panels. In the left panels, the dashed white lines represent the initial parabolic orbit of the star (top panel) and the Keplerian elliptical orbit of the most bound debris (middle and bottom panels); white solid circles demarcate spheres with radii equal to $r_{\rm t}$, $a_{\rm mb}$, and $r_{\rm a}$ . White dots connected by solid lines indicate the position of the photospheric radii for observers in the orbital plane. The photosphere is in an outflow of low-density material, which radially expands and becomes quasi-spherical around the stream's orbital pericenter, where the majority of dissipation occurs.}
    \label{fig:denproj}
\end{figure*}
In contrast to the earlier TDE simulations of \citet{SS24}, we used an updated version of \texttt{RICH}, and we summarize the main adjustments here. The diffusion coefficient for radiation transport is $D\equiv c\lambda/\alpha_\text{Ross}$, where $c$ is the speed of light and $\lambda$ is now\footnote{Previously, \citet{SS24} used the Larsen flux limiter.} the dimensionless flux-limiter from \citet{Krum07}:
\begin{equation}
\label{eq: diff coeff}
    \lambda = \frac{1}{\xi}(\coth{\xi}-\frac{1}{\xi}).
\end{equation}
This expression provides a smooth transition between the optically thin and the optically thick regimes, ensuring that the radiation flux remains physical in all conditions\footnote{In optically thick regions $\xi \to 0$, $\lambda \to \frac{1}{3}$ and we recover the standard diffusion limit, while in optically thin regions $\xi \to \infty$, $\lambda \sim \frac{1}{\xi}$ and we approach the free-streaming limit.}. The parameter $\xi^{-1}$ is the length scale of the problem (in units of the mean free path) for the variation in the radiation energy density, defined as
\begin{equation}
\label{eq: grad}
    \xi = \frac{|\nabla u_\text{rad}|}{\alpha_\text{Ross}u_\text{rad}},
\end{equation}
where $u_\text{rad}$ is the radiation energy density, $\nabla u_\text{rad}$ its gradient (see Appendix \ref{app: grad}) and $\alpha_\text{Ross}$ is the Rosseland mean absorption coefficient (units of ${\rm cm}^{-1}$). The latter is a function of the gas temperature $T$ and the density $\rho$ obtained as a weighted mean over frequencies $\nu$:
\begin{equation}
    \frac{1}{\alpha_\text{Ross}} = \frac{\int_0^\infty [(\alpha_\nu+\sigma_\nu)^{-1}\partial B_\nu/\partial T] {\rm d}\nu}{\int_0^\infty[\partial B_\nu/\partial T] {\rm d}\nu},
\end{equation}
where $\alpha_\nu=\alpha_\nu(T,\rho)$ and $\sigma_\nu=\sigma_\nu(T,\rho)$ are the frequency-dependent absorption and scattering coefficients, and $B_\nu =B_\nu(T)$ is the Planck function. For opacities, we used tabulated values (see Appendix \ref{app: opacity} for further details);
we follow \citet{Krief16} and assume Solar abundances and local thermodynamic equilibrium, which are more appropriate for high density material than the tables from CLOUDY \citep{CLOUDY17} used in \citet{SS24}.
Compton cooling is dynamically accounted for during the simulation under a grey approximation, i.e., assuming the radiation field to be Planckian with a temperature $T_{\rm{rad}}$ such that $aT_{\rm{rad}}^4=u_{\rm{rad}}$ (where $a=4\sigma_{\rm{SB}}/c$, with $\sigma_{\rm{SB}}$ being the Stefan–Boltzmann constant).  
We used an initial number of cells $N = 4\times10^4\sqrt{M_{BH}/M_\odot}=4\times10^6$.
At each time-step, cells at distance $r$ from the IMBH are refined\footnote{The refinement does not apply to cells at $r<1.75r_\text{t}$ or $r>6a_\text{mb}$.} if any of the following conditions is satisfied:
\begin{enumerate}[(i)]
    \item Their mass is larger than a maximum, space-time-dependent threshold, which is 
    \begin{equation}
    M_{\text{max}}(r,t) =\min[1, (0.05\cdot r/r_\text{t})^{2.5})]
    \begin{cases}
        \mathcal{M}_{\rm max} \quad \text{if} \quad x\geq-9a_{\rm mb} \\
        30\mathcal{M}_{\rm max} \quad \text{otherwise}, 
    \end{cases}
    \end{equation}
    where $\mathcal{M}_{\rm{max}}\equiv 3.75\cdot10^{-8}M_\star$ and $a_\text{mb}=r_\text{t}^2/(2r_\star)$ is the semi-major axis of the most bound debris. 
    This scheme decreases mass resolution as a function of distance in order to increase the resolution near pericenter;
    \item Their volume is larger than $10^{-5}$ times the volume of the simulation domain;
    \item Their volume is larger than $V_{\rm p}\left(r/r_\text{t}\right)^{1.5}$ and their density is larger than $0.01\rho_{\rm X}$. The volume $V_{\rm p}\equiv 4\pi \left(r_{\rm t}/125\right)^3/3$ is the target volume of a cell at pericenter and $\rho_{\rm X}$ is calculated as the maximum density of the gas with $x > 0.8r_{\rm t}$. The spatial scaling of $1.5$ results from the fact that the width of the outgoing stream generally scales as $\propto \sqrt{r}$ \citep{Coughlin16}.
\end{enumerate}
In contrast, cells are removed if all the following conditions are satisfied:
\begin{enumerate}[(i)]
    \item Their mass is lower than a minimum space-time-dependent threshold, which is
    \begin{equation}
    M_{\text{min}}(r,t) =\min[1, (0.05\cdot r/r_\text{t})^{2.5})]
    \begin{cases}
        \mathcal{M}_{\text{min}} \quad \text{if} \quad x\geq-5a_\text{mb} \\ 
        30\mathcal{M}_{\text{min}} \quad \text{otherwise}, 
    \end{cases}
    \end{equation}
    where $\mathcal{M}_{\text{min}}\equiv 8.75\cdot10^{-9}M_\star$.
    \item Their volume is less than $5\cdot10^{-6}$ the domain size or smaller by a factor of 2.5 than the one of the neighbour cells.
    \item Their density is lower than $\rho_{\rm X}$ or their volume is smaller than $0.2V_{\rm p}\left(r/r_{\rm t}\right)^{1.5}$.
    \item If the volume of the cell is less than $0.025V_{\rm p}$. In this case, the cell is always removed.
\end{enumerate}
\begin{figure}
    \centering
    \includegraphics[width=\linewidth]{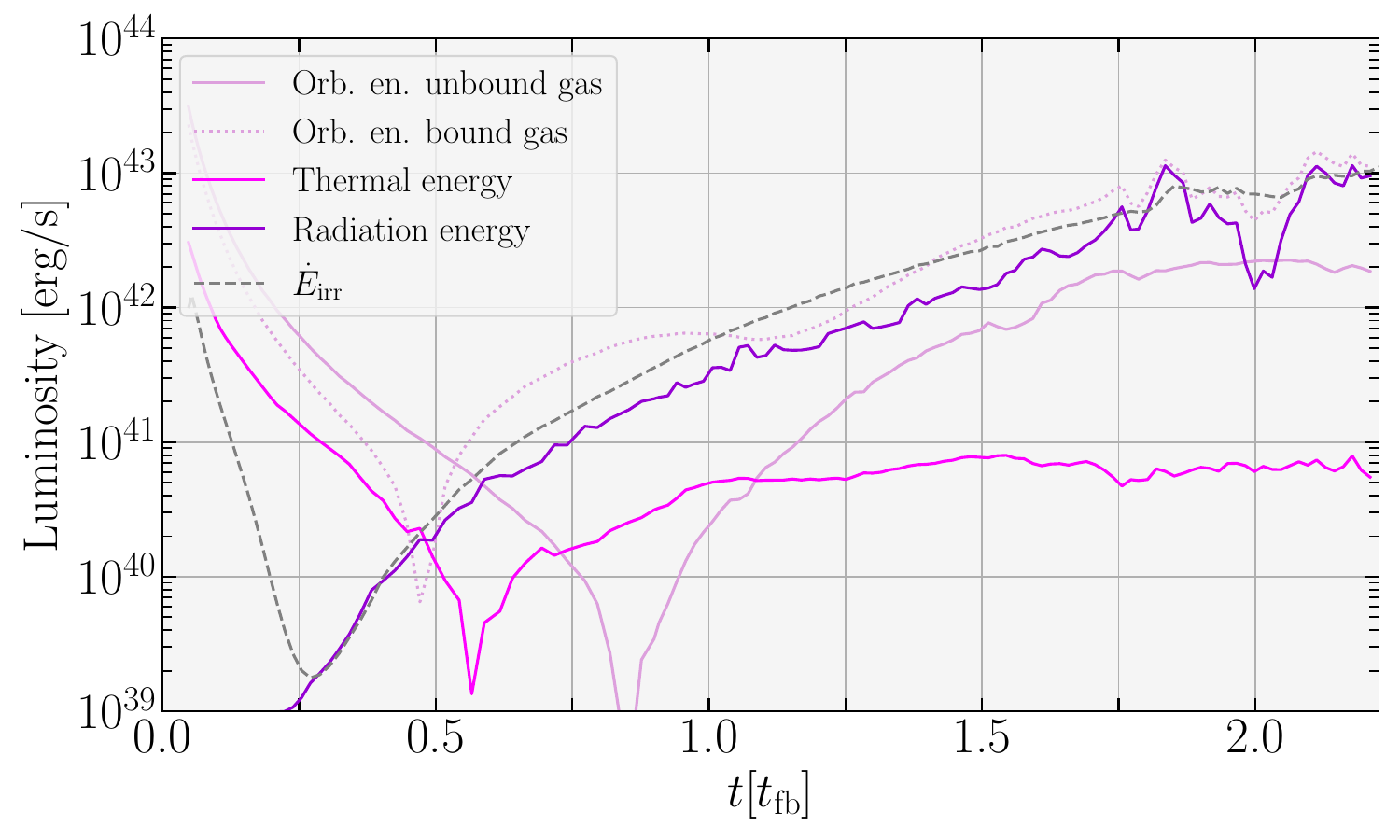}
    \caption{Luminosities from orbital energy of unbound (solid light pink line) and bound gas (dotted light pink line), thermal energy (magenta line), radiation energy (purple line), and dissipation energy (dashed gray line), plotted against time $t$ normalized to the fallback time $t_{\rm fb}$. After $0.5t_{\rm fb}$ dissipation rate tracks the evolution of radiation energy.}
    \label{fig:ratesE}
\end{figure}
\begin{figure*}
    \centering
    \includegraphics[width=
    \linewidth]{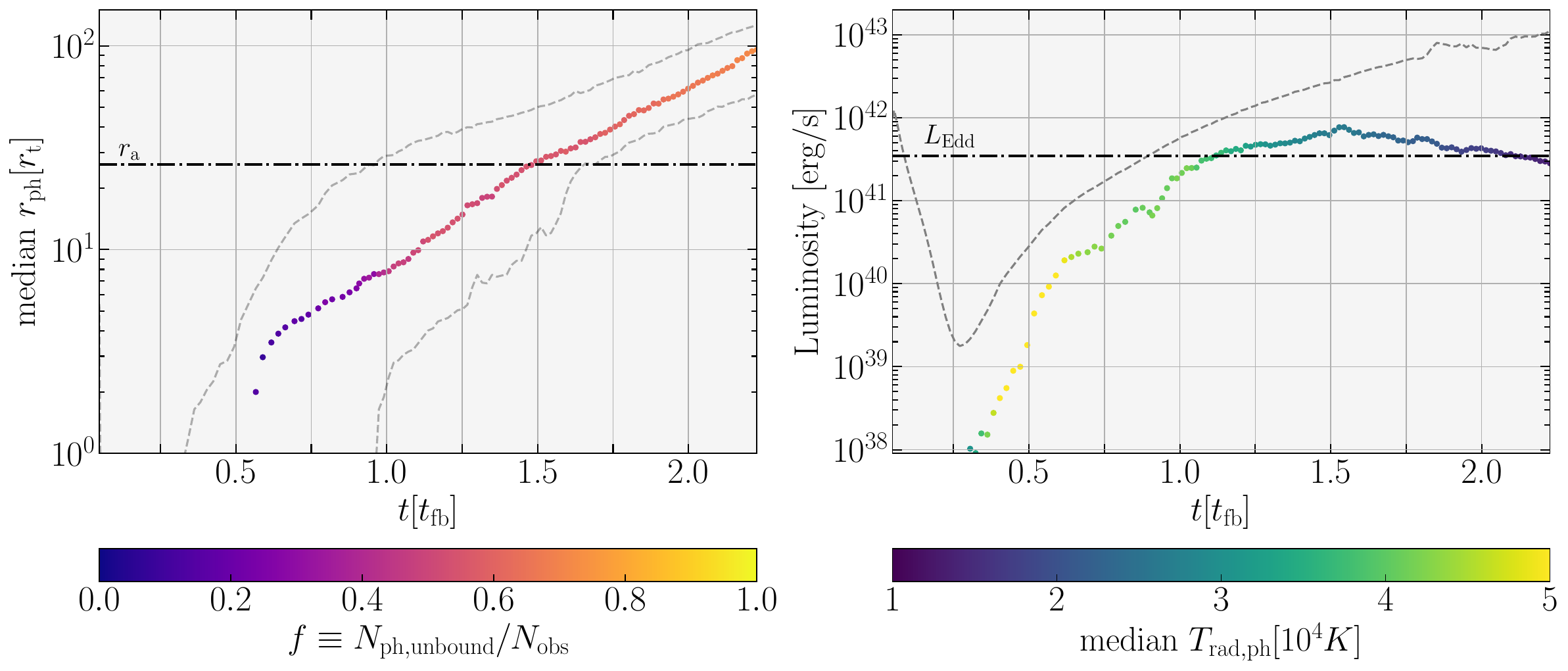}
    \caption{\emph{Left panel}: time evolution of the photospheric radius, $r_{\rm ph}$, computed as a median value over different lines of sight. Dashed gray lines define $1\sigma$ scatter around the median among the total number of evenly spaced sight lines, $N_{\rm obs} = 192$. Points are coloured by the fraction $f$ of unbound photospheric cells. The photosphere expands beyond the apocenter of the most bound debris ($r_{\rm a}$, the dash-dotted black line) and becomes more isotropic and unbound. \emph{Right panel}: FLD bolometric light curve. Points are coloured with the radiation temperature at the photospheric radius (computed as a median across 192 sight lines). The dashed gray line shows the total volume-integrated dissipation rate $\dot{E}_{\rm irr}$.  
    By the end of the simulation, both the dissipation rate and the emergent luminosity overcome the Eddington limit $L_{\rm Edd}$ (dashed-dotted black line), though the dissipation rate exceeds the FLD luminosity by over one order of magnitude. The FLD luminosity peaks at roughly $\approx 2 L_{\rm Edd}$ before declining and then levelling at $L_{\rm Edd}$.  The shock dissipation rate is orders of magnitude below naive estimates for radiatively efficient accretion of fallback material, $\sim 0.1 \dot{M}_{\rm fb} c^2 \sim 7 \times 10^{46}~{\rm erg~s}^{-1}$.} 
    \label{fig:fld}
\end{figure*}
Precise implementation details of the moving mesh refinement scheme can be found in the open source \texttt{RICH} repository\footnote{\url{https://gitlab.com/eladtan/RICH}}. 
The simulation with the described resolution scheme is our {\it fiducial} run, which we dub \emph{High res} run. In order to assess the robustness of our results, we repeat the same simulation twice with a lower resolution, hereafter called \emph{Low res} and \emph{Middle res}. These non-fiducial runs are performed by decreasing the initial number of cells $N$ by a factor of sixteen and four respectively, and by adjusting accordingly the minimum cell size and maximum cell mass thresholds by the same factors. We run the fiducial simulation from the disruption of the star up to $t \approx 2.2 ~t_{\rm fb}$, where $t_{\rm fb}= 4 \,\text{days}\,(M_\text{BH}/(10^4M_\odot))^{1/2}(M_\star/M_\odot)^{-1} (r_\star/R_\odot)^{3/2}\approx 2.5 \,\text{days}$ is the approximate fallback time \citep{stone13} of the most bound debris\footnote{In reality, $t_{\rm fb}$ is an analytic approximation, and a very low-mass tail of the stellar debris makes its first return to pericenter much earlier, at $\approx 0.1 t_{\rm fb}$.}. The \emph{Low} and \emph{Middle res} runs are executed up to $\approx1.5~t_\text{fb}$ and $\approx2.5~t_\text{fb}$, respectively. 
We achieve unprecedented resolution in global 3D (end-to-end) simulations of TDEs, as shown in the cumulative distributions in Fig.\ref{fig:hist} and summarized in Table \ref{tab: res}. At $t\approx1.5t_{\rm fb}$, cell masses and sizes reach a minimum of, respectively, $10^{-15}M_\odot$ and $0.04R_\odot$ in the {\it High res}, in the orbital plane. We compare our range of cell masses and sizes to typical resolutions adopted in previous studies (vertical lines in Fig.\ref{fig:hist}), reporting typical cell sizes for grid-based simulations and typical particle masses for SPH simulations.  Our typical cell masses and sizes are generally smaller than those achieved in other end-to-end simulations \citep{Ryu23, Price24}, although the peak resolution of \citet{Ryu23} is higher than our peak resolution by a factor $\approx 2$.  The resolution in this paper is also significantly higher than in typical stream injection simulations \citep{BonnerotLu20, Huang24} as well as more short-term simulations focused on debris properties \citep{Norman21, Fancher23}.  Recent SPH simulations focused on achieving convergence of the nozzle shock \citep{Hu25, Kubli25} obtain peak mass resolutions comparable to the high-resolution tail of our cell masses.

\section{Results}
\label{sec: res}
In this section, we present our results, primarily showing the \emph{High res} data, along with convergence tests to evaluate the sensitivity of our results to resolution.
Figure \ref{fig:denproj} shows snapshots at three representative times, illustrating the projection (see Appendix \ref{app: denproj} for further technical explanation) on the orbital plane of the mass density (left column) and the dissipation energy density rate (right column). 
The finite size of the star spreads its debris across a range of specific orbital energies: $\Delta\varepsilon\approx GM_\text{BH}r_\star/r_\text{t}^2$ \citep{stone13}, with each fluid element of the disrupted star flying quasi-ballistically away from pericenter, but on orbits that differ from the initial (nearly parabolic) one.
The bound half of the star returns to the IMBH, and during its first pericenter return it is dramatically compressed in the vertical direction, dissipating orbital energy through the nozzle shock \citep{Guillochon14, BonnerotStone21}. Other shocks ensue at later times as tightly bound, post-pericenter debris interacts with fresh stellar material returning for the first time \citep{Hayasaki13, Shio15, Piran15}. While apsidal precession could amplify these interactions, it is negligible for our choice of parameters.
Orbital energy is converted in these shocks into thermal and radiation energy, which  is transported outward and (after some adiabatic losses) eventually leaves as radiation through the photosphere, as will be quantified in the next subsection. The photosphere resides within a low-density outflow that expands in time radially and becomes more spherical around the pericenter region, as can be seen from the white dots in the left column of Fig. \ref{fig:denproj}. 
Just after disruption, most of the energy dissipation occurs within the bulk of the stellar debris.
As the event progresses, the nozzle shock becomes the main site of dissipation, yet its efficiency remains too low to trigger disc formation, allowing much of the debris to persist on highly eccentric orbits.

\subsection{Energy dissipation}
\label{sec:diss}
In TDEs, care must be taken to differentiate between dissipation (i.e. energy change due to irreversible work) and reversible increases in gas energy due to compression of returning streams.  
Particularly at early times, heating from adiabatic compression can dominate over heating from the nozzle shock (see e.g. Fig. 11 in \citealt{Andal22}), so in Figs \ref{fig:ratesE} and \ref{fig:fld} we show the dissipation rate $\dot{E}_{\rm irr}$ (dashed gray line). This quantity is computed in \texttt{RICH} during each hydro numerical step as the difference between the total energy change and the portion associated with reversible work as:
\begin{equation}
\label{eq: uirr}
     \dot{E}_{\rm irr} \equiv  - \sum_{\text{interfaces}} A (P_*\vec{v_*}
     -\vec{v}_{\rm c}P_*
     -P_{\rm c}\vec{v_*}
     )\cdot \vec{n}_{\rm A},
\end{equation}
where $A$ is the area of the interface (and $\vec{n}_{\rm A}$ its normal vector), $P_{\rm c},\vec{v}_{\rm c}$ are gas pressure and velocity at the cell centre and the starred quantities are computed at the contact discontinuity through Riemann solvers (see Appendix \ref{app: diss} for further details). The second and third term on the right-hand side of Eq.\eqref{eq: uirr} correspond to the change in, respectively, orbital and thermal energy. The latter is transformed, in the radiation numerical step, in radiation energy.
Figure \ref{fig:ratesE} show the evolution of luminosities associated with orbital, thermal, and radiation energy, computed as the absolute value of the net energy rate across the entire simulation volume.
After an initial peak associated with the stellar disruption, the dissipation rate declines by nearly three orders of magnitude as the system relaxes. For $t\geq0.25t_{\rm fb}$, dissipation begin to rise again due to the nozzle shock. The irreversible work primarily increases the gas thermal energy, whereas after about $0.5t_{\rm fb}$ the dissipation rate tracks the evolution of radiation energy.

In analytic model-building, approximate dissipation rates are often computed using the mass fallback rate $\dot{M}_{\rm fb}(t)$ \citep{Loeb97,Strubbe09}, which we show as a dashed black line in Fig.~\ref{fig:MdotW}. It can be described as the fallback rate of the 
bound debris under Keplerian motion \citep[e.g.,][]{Evans89}:
\begin{equation} 
\label{eq:Mdot}
   \dot{\rm{M}}_{\rm{fb}}(t) = \frac{{\rm dM}}{{\rm d}\varepsilon} \frac{{\rm d}\varepsilon}{{\rm d}t}=
   -\frac{1}{3}\frac{{\rm d}\rm{M}}{{\rm d}\varepsilon}(2\pi G M_{\rm BH})^{2/3}t^{-5/3}, 
\end{equation}
where ${\rm d}\rm{M}/{\rm d}\varepsilon$ is the frozen-in orbital energy spread, which we computed numerically immediately after ($0.05~t_{\rm fb}$) disruption. Past work often makes the simplifying assumption that the dissipation rate and emergent luminosity can both be written as $L_{\rm fb}(t) = \eta \dot{M}_{\rm fb}(t) c^2$, where $\eta$ is a dimensionless constant.  If we take $\eta=0.1$ for simplicity\footnote{$\eta\sim 0.1$ is consistent with radiatively efficient accretion, though see Sec.\ref{sec:compar_sim} for further comparisons}, then the peak luminosity $L_{\rm fb} \sim 7 \times 10^{46}~{\rm erg~s}^{-1}$ is over 3 orders of magnitude larger than the maximum dissipation rate visible in Fig. \ref{fig:fld}.  It is therefore clear that shocks are dissipating only a very small fraction of the debris orbital energy in our simulation, further corroborating evidence against efficient circularization (see Sec. \ref{sec: ecc}).

\subsection{Luminosity and photosphere}
\label{sec:rad}
As \texttt{RICH} self-consistently evolves the radiation field, we extract the bolometric light curve reproducing the FLD scheme implemented in our simulations. First we find the photosphere, i.e., the surface at which photons have equal probability either to be retained or to escape. We discretize the 3D space with \texttt{HEALPix} \citep{healpix}, creating $N_\text{obs}=192$ observers spacing the same solid angle. For each of them, we defined the photospheric radius as $r_{\rm ph}$ such that
\begin{equation}
    \tau(r_{\rm{ph}})\equiv\int^\infty_{r_\text{ph}}\alpha_\text{Ross} {\rm d}r= \frac{2}{3}, 
\end{equation} 
where $\tau$ is the optical depth and the integration is performed radially inward. The left panel of Fig. \ref{fig:fld} illustrates the temporal evolution of the photosphere, specifically the median value of $r_{\rm ph}$ along the 192 lines of sight.  A finite median first appears at approximately $0.5t_{\rm fb}$ and then expands, reaching a final (median) radius of $\approx 100 r_{\rm t} $, few times $r_{\rm a}$, the apocenter of the most tightly bound debris. Initially confined to the orbital plane and exhibiting strong asymmetry, the photospheric surface  becomes increasingly spherical, as can be noticed from the thinning of the dispersion (dashed lines in left panel in Fig. \ref{fig:fld}).

Assuming grey radiation and flux-limited diffusion, we computed the flux $F$ through its photosphere as in \citet{Krum07}:
\begin{equation}
    F = -\frac{c\lambda}{\alpha_\text{Ross}}\nabla_{\rm r} u_\text{rad} \label{eq: F results fld}, 
\end{equation}
where $\nabla_r u_\text{rad}$ is the projection of the gradient of radiation energy density along the radial direction.
The parameter $\lambda$ is the dimensionless flux-limiter given by Eq.\eqref{eq: diff coeff} and the opacity is treated as described in Sec. \ref{sec: meth} and Appendix \ref{app: opacity}. 
Finally, the bolometric luminosity is found by averaging over the observers\footnote{We checked that increasing the number of observers does not affect the results for the FLD curve.}:
\begin{equation}
\label{eq:L fld}
    L_\text{FLD}=\frac{1}{N_\text{obs}}\sum_{i=1}^{N_\text{obs}} L_i = \frac{4\pi}{N_\text{obs}}\sum_{i=1}^{N_\text{obs}}r_{\text{ph}_i}^2F_i,
\end{equation}
with $F_i$ given by\footnote{In the rare cases where the flux obtained with Eq.\eqref{eq: F results fld} for an observer $i$ is negative or makes the luminosity overcome the maximum value ($L_{\text{max},i} =  4\pi c u_{\text{rad},i}r_{\text{ph},i}^2 $), we used the latter as value for luminosity i.e. $L_i=L_{\text{max},i}$.}  Eq. \eqref{eq: F results fld}.
As shown in the right panel of Fig. \ref{fig:fld}, the light curve peaks around  $t_{\rm p}\approx1.5t_\text{fb}$, overcoming the Eddington luminosity $L_\text{Edd}\equiv 4\pi G M_{\rm BH}c/\kappa_{\rm p} \approx3\times 10^{41}$ erg/s, with $\kappa_{\rm p}=\alpha_{\rm Ross}(t_{\rm p})/\rho(t_{\rm p})$ being the average value of the opacity at the time of the peak. We underline that $\kappa_{\rm p}$ is computed self-consistently from the opacities in our simulation as $\kappa_{\rm p} = \big(\sum_{i\leq N_{\rm obs}}1/\kappa_i\big)^{-1}$, which gives greater weight to optically thin lines of sight \textendash the directions from which most of the emergent radiation reaches the observer. This yields $\kappa_{\rm p}\approx1.4$, a few times larger than the Thomson scattering opacity typically assumed. 
The median radiation temperature measured at the photosphere is $\sim$ a few $\times 10^4$ K. 
At late times, the light curve plateaus at $L_{\rm Edd}$, while  the total dissipation rate (dashed line) is larger by one order of magnitude, indicating that an optically thick region enveloping the shock dissipation zones prevents most of the dissipated orbital energy from being promptly radiated. 

\subsection{Hydrodynamics: eccentric disc plus wind}
\label{sec:hydro}
\subsubsection{Flow eccentricity}
\label{sec: ecc}
\begin{figure}
    \centering    \includegraphics[width=\linewidth]{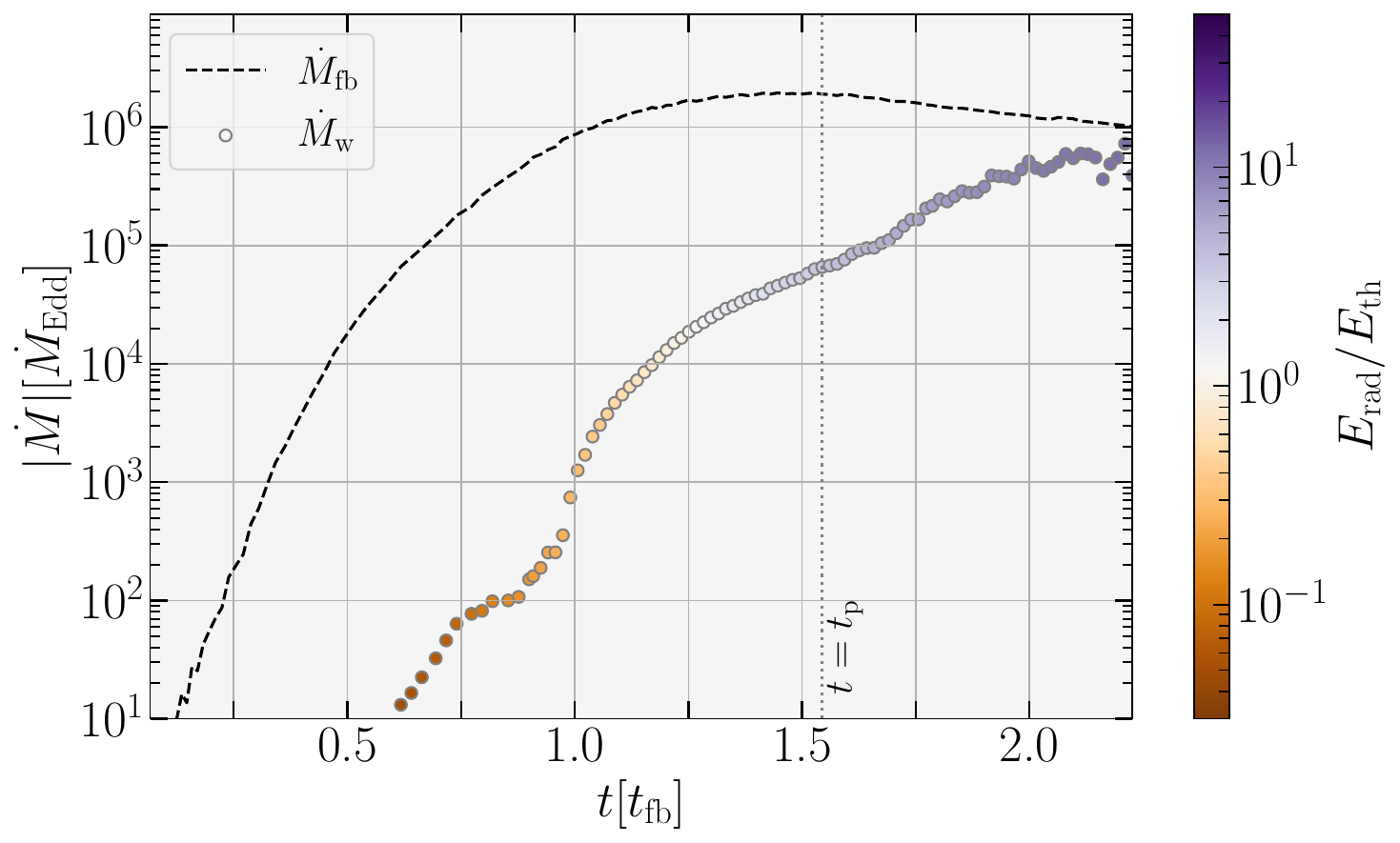}
    \caption{Time evolution of the wind mass rate ($\dot{M}_{\rm w}$, scatter plot) at $r=0.5a_{\rm mb}$, colour-coded by the ratio of volume-integrated (over the whole wind material) radiation energy $E_{\rm rad}$ and gas thermal energy $E_{\rm th}$.
    A super-Eddington fallback rate ($\dot{M}_{\rm fb}$, dashed black line) drives a super-Eddington outflow that becomes radiation-dominated after $\approx1.5~t_{\rm fb}$, where a naive Eddington rate is defined as $\dot{M}_{\rm Edd}=L_{\rm Edd}/c^2$. The vertical dotted line shows the time $t_{\rm p}$ of the peak of the light curve.  The rate of mass outflow in the wind is always a small but not negligible fraction of $\dot{M}_{\rm fb}$, with an efficiency $\zeta\equiv |\dot{M}_{\rm w}/\dot{M}_{\rm fb}|\approx0.5$ by the end of the simulation.}
    \label{fig:MdotW}
\end{figure}
\begin{figure}
    \centering
    \includegraphics[width=\linewidth]{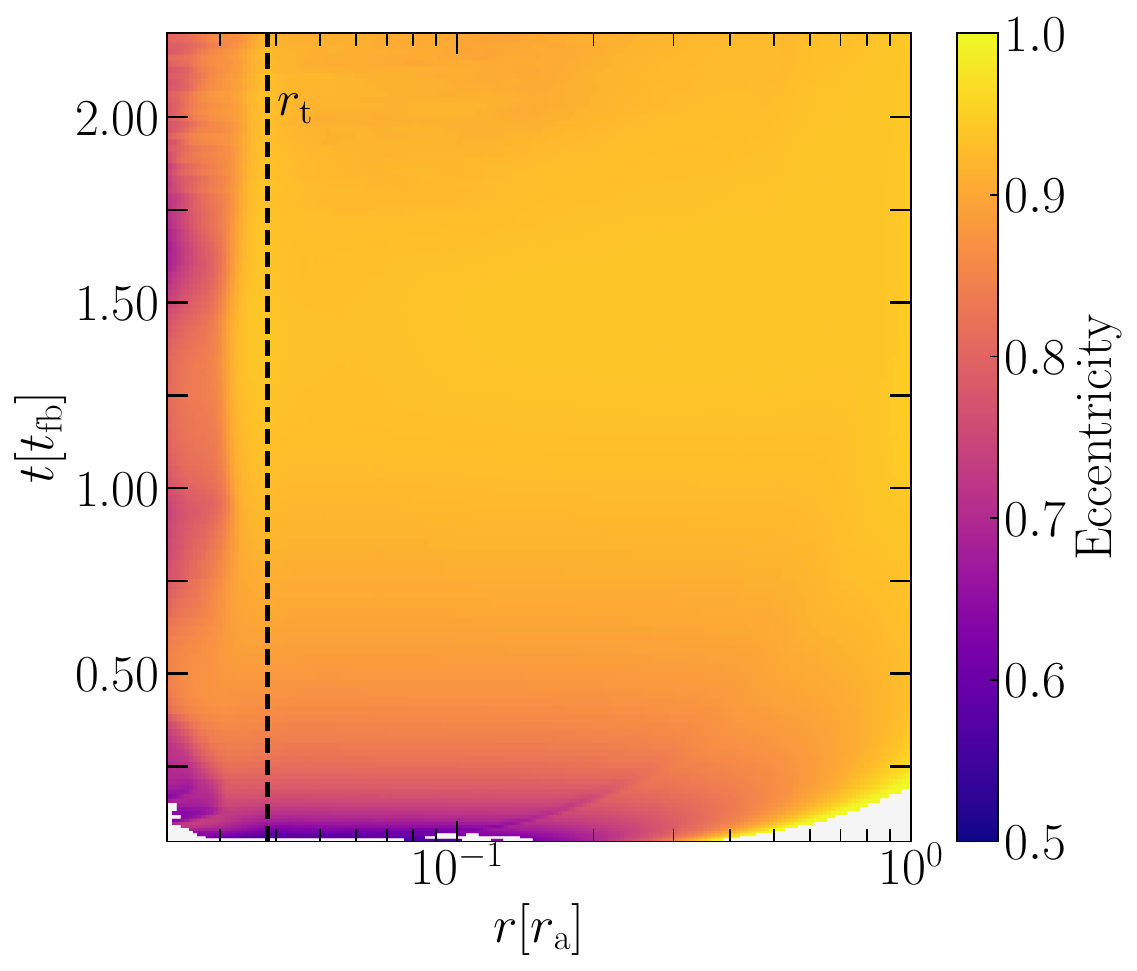}
    \caption{Time evolution of the bound gas eccentricity as a function of radial distance (from $r_0$ to $r_\text{a}$) from the IMBH. Each point is a mass-weighted average on spherical shells at constant $t$. The vertical dashed line shows the tidal radius. Over the course of the simulation, very little circularization occurs, unlike in a related simulation of a SMBH TDE \citep{SS24}.  The lack of debris circularization evident here aligns with the visual appearance of the debris in Fig. \ref{fig:denproj} and the low dissipation rate seen in Fig. \ref{fig:ratesE}.}
    \label{fig:ecc}
\end{figure}
Visual inspection of Fig. \ref{fig:denproj} indicates that little circularisation of the flow has taken place over the investigated timescale, and thus there is no formation of a classical accretion disc. To quantitatively assess this observation, we computed the gas eccentricity $e$ of all material with negative orbital energy:
\begin{equation}
\label{eq:ecc}
    e = \sqrt{1 + \frac{2\varepsilon j^2}{(GM_{\text{BH}})^2}},
\end{equation}
where $\varepsilon=0.5|{\bf{v}}|^2-GM_\text{BH}/(r-2r_\text{g})$ is the specific orbital energy and $j = |\bf{j}|=|\bf{r}\times {\bf{v}}|$ the magnitude of the specific angular momentum, with $\bf{v}$ being the velocity vector. 
Fig. \ref{fig:ecc} shows the space-time evolution of this eccentricity. Each point in the diagram is obtained as a mass-weighted average on spherical shells, 
\begin{equation}
    e(r,t) = \frac{\Sigma_{i} e_i m_i}{\Sigma_{i}m_i},
\end{equation}
where $e_i$ is given by Eq.\eqref{eq:ecc}. The eccentricity is lower in the inner regions, reaching a minimum of $e\approx 0.7$ inside $r_{\rm t}$, but overall the material has not circularized yet.  At most times, typical bound debris retain an eccentricity  comparable to the ``frozen-in'' eccentricity of the most bound debris, $e_{\rm mb}=1-2r_\star/(\beta r_\text{t})\approx 0.93$. 

\subsubsection{The outflow}
\label{sec: wind}
\begin{figure}
    \centering
    \includegraphics[width=.85\linewidth]{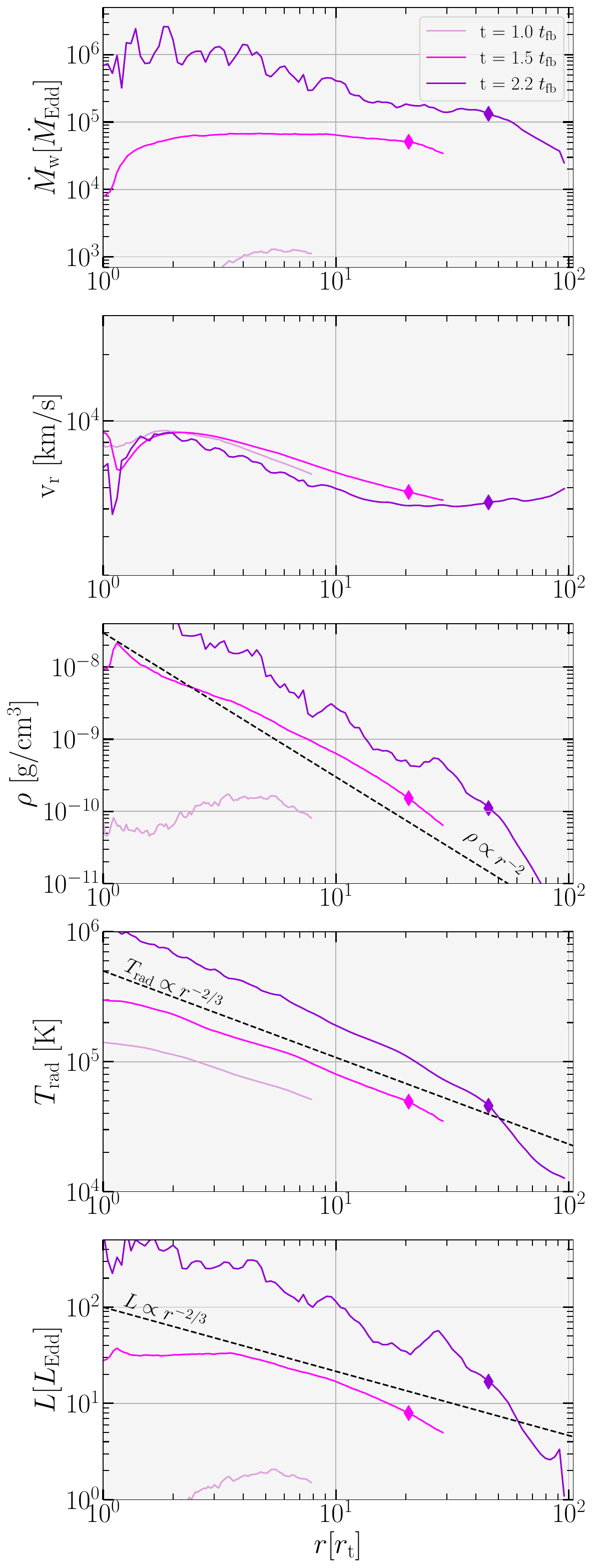}
    \caption{Radial profiles at $t = 1 t_{\rm fb}$ (light pink line), $t_{\rm p} = 1.5t_{\rm fb}$ (magenta line) and the last time available, $t = 2.2t_{\rm fb}$ (purple line). Solid coloured lines are obtained as mass-weighted (except for the wind mass rate) means on spherical shells and span from the tidal radius up to the median photospheric radius. Diamonds on the lines mark the median trapping radius among the lines of sight that contain an advective regime. 
    \emph{First panel}: Wind mass rate profiles. \emph{Second panel}: Radial velocity profiles. \emph{Third panel}: Density profiles. The dashed line shows the analytically predicted power-law $\rho\propto r^{-2}$. \emph{Fourth panel}: Radiation temperature profile. The dashed line shows the analytically predicted power-law $T_{\rm rad}\propto u_{\rm rad}^{1/4}\propto r^{-2/3}$. \emph{Fifth panel}: Advection luminosity profile. The dashed line shows the analytically predicted power-law $L\propto r^{-2/3}$. After $t = 1.5t_{\rm fb}$, the outflow behaves as an adiabatic radiative driven wind.}
    \label{fig:den prof}
\end{figure}
Quantitative analysis of the large-scale debris ($r>r_{\rm a}$) visible in the last panel of Fig.\ref{fig:denproj} shows that this material is outflowing (i.e., it has a positive radial velocity $v_{\rm r}$), encompassing the photosphere. 
In contrast to \citet{Price24}, the expanding material in our simulation becomes unbound, forming a wind rather than a bound envelope. We identify material as unbound when its Bernoulli number $\mathcal{B}\equiv\varepsilon+\varepsilon_{\rm th}+(P+u_{\rm rad}/3)/\rho$, is positive, where $\varepsilon_{\rm th}$ and $P$ are respectively the gas specific thermal energy and pressure , and $u_{\rm rad}/3$ is radiation pressure. The colour of the points in the left panel of Fig.\ref{fig:fld} indicates the ratio $f\equiv N_{\rm ph, obs}/N_{\rm obs}$, which measures the fraction of lines of sight for which the outflowing material is unbound at the photosphere relative to the total number $N_{\rm obs}$ . We note that this ratio rises\footnote{This behaviour is consistent across all resolutions (see, Fig. \ref{fig:f_conv}).} to $\approx70\%$ by the end of the simulation, indicating that most lines of sight intercept unbound material. 
A radiation-dominated outflow is expected given the super-Eddington dissipation rate, evident in Fig. \ref{fig:fld}: when rates of energy production exceed $L_{\rm Edd}$, astrophysical systems usually begin to release large fractions of their energy production in the form of kinetic luminosity \citep{Blandford99, Ohsuga11, Jiang14}. As described in Sec.\ref{sec:diss},  IMBH TDEs  appear to follow this general trend. We quantitatively show in this section that this is indeed the case.

Fig.~\ref{fig:MdotW} shows the wind mass rate (scatter line) through a sphere of radius $\bar{r}=0.5a_{\rm mb}$. We approximated it as:
\begin{equation}
 \dot{M}_\text{\rm w}(r,t)= C(t)\pi\sum_{i:r_i= \bar{r}} s_i^2(t)\rho_i(t) v_{{\rm r},i}(t) \quad \text{with} \quad C(t)\equiv \frac{4\pi r^2}{\pi\sum_i  s_i^2(t)}
\end{equation}
where $C(t)$ is a normalisation factor taking into account cells selection, $s_i=V_i^{1/3}$ is the size of the $i-$th cell of volume $V_i$, and the sum is done only considering wind (i.e., outflowing and unbound) cells. Although the wind represents only a small fraction of the fallback rate for most of the evolution ($\lesssim10\%$ until $1.75~t_{\rm fb}$), its mass-loss rate grows and reaches $\dot{M}_{\rm{w}}(t)\approx10^6\dot{M}_{\rm Edd}$, corresponding to roughly half of $\dot{M}_{\rm fb}$, at the end of the simulation, where $\dot{M}_{\rm Edd} \equiv L_{\rm Edd}/c^2$  is the Eddington mass rate in maximally efficient conditions.
In this super Eddington outflow, radiation energy dominates over the thermal energy of the gas after approximately $1.5~t_{\rm fb}$,  and at the end of the simulation the energy ratio is close to $\sim 10$, as visible by the colours of the points in Fig.\ref{fig:MdotW}.
\begin{figure*}
    \centering
    \includegraphics[width=\linewidth]{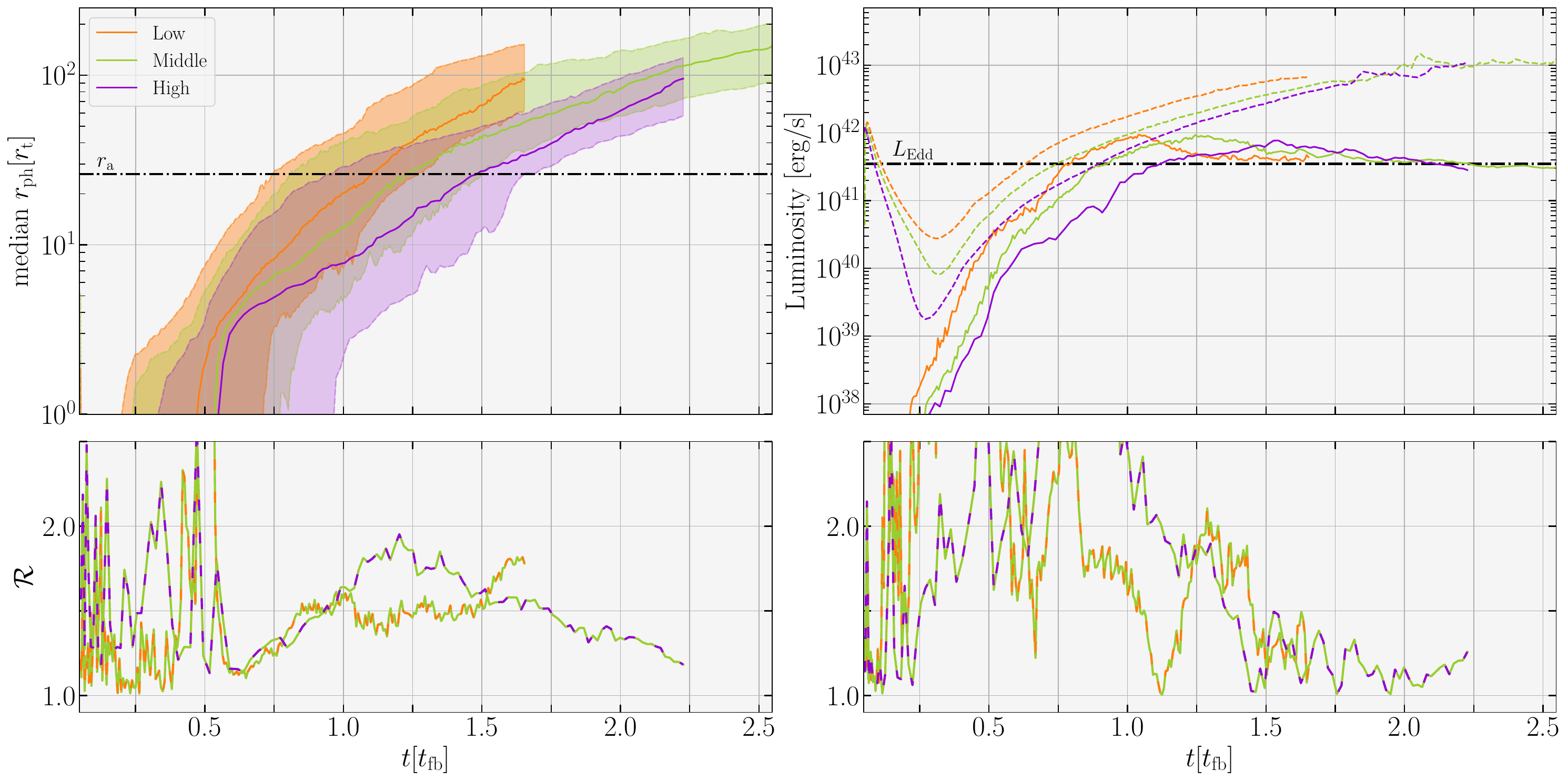}
    \caption{Resolution tests for radiation. Orange, green and purple lines are, respectively, for \emph{Low, Middle} and \emph{High res}. \emph{Upper left panel}: Time evolution of the median value of the photospheric radii $r_{\rm ph}$. As $r_{\rm ph}$ varies depending on line-of-sight, the shaded regions show $1\sigma$ contours across an isotropically distributed set of viewing directions. \emph{Upper right panel}: light curves (solid lines) according to the FLD scheme, and dissipation rates (dashed lines). \emph{Lower panels}: Time evolution of $\mathcal{R}$ between resolutions for photosphere ({\it left}) and light curve ({\it right}). The dashed orange-green  and dashed violet-green lines compare, respectively, \emph{Middle} and \emph{Low res} and \emph{Middle} and \emph{High res}. Resolutions converge by a factor of $\lesssim2$. Disregarding the resolution, the photosphere expands in time, and the light curve overcomes the Eddington limit and settles to that value at later times.}
    \label{fig:fld_R_conv}
\end{figure*}

Fig. \ref{fig:den prof} shows radial profiles of the main properties of the wind, at different times. With the except for $\dot{M}_{\rm w}$ in the first panel, quantities are computed as mass-weighted means on spherical shells\footnote{We note that the wind is symmetric above and below the orbital plane, while it presents an anisotropic configuration otherwise, but a more detailed analysis is left to a future work.}, from the tidal radius out to the median photospheric radius.  While the mass flow rate and the radial velocity remain approximately constant, decreasing by at most a factor of a few across this radial distance, density, radiation temperature, and advection luminosity fall by multiple orders of magnitude as approximate power laws.
In order to validate and interpret our findings in Fig. \ref{fig:den prof}, we follow the approach outlined by \citet{Rossi09, Linial24} and construct a simple model for an optically thick (see, Sec.\ref{sec:rad}), radiation driven (Fig.~\ref{fig:MdotW}) wind, with a mass outflow rate $\dot{M}_{\rm w}$, launched at $r=r_{\rm diss}$. 
The total energy per unit time available from the returning debris, $L (r_{\rm diss}) = GM\dot{M_{\rm w}}/r_{\rm diss}$, is deposited mainly in radiation at the base of the wind. At small radii, this radiation energy is transported outwards by advection, as the high optical depth implies diffusion timescales, $t_\text{diff}\equiv r \tau/c$, longer than dynamical ones ($t_{\rm dyn} \equiv r/v_{\rm r}$). As the wind expands radially, the mass density  $  \rho(r, t) = \dot{M}_{\rm w}(t)/(4\pi r^2 v_{\rm r})$ decreases, as does the optical depth $\tau \sim \kappa \rho r$, until eventually the two timescales become equal at the trapping radius $r_{\rm tr} = c/(v_{\rm r} \kappa \rho)$.  This radius is marked with diamonds in the profiles shown in Fig.~\ref{fig:den prof}. Since in our simulation the times taken by the wind to reach the trapping radius are shorter than the time scale over which the mass loss rate changes, i.e., $t_{\rm dyn}(r_{\rm tr}) \ll t_{\rm fb}$, $\dot{M}_{\rm w}(t)$ is roughly constant in radius ($\dot{M}_{\rm w}(r,t) \approx \dot{M}_{\rm w}(t)$) and $  \rho(r, t) \propto r^{-2}$ (see, Fig. \ref{fig:den prof}). 
The radial extent of the advection dominated region depends on how super-Eddington the mass flow rate is, since $\dot{M}_{\rm w}(t)/\dot{M}_{\rm Edd}=r_{\rm tr}/r_{\rm g}$; in our simulation, $r_{\rm tr}$ increases with time because the mass loss rate becomes more vigorous as time progresses (see, Fig.~\ref{fig:den prof} ).  For $r \lesssim r_{\rm tr}$, the radiation energy density decreases adiabatically with radius as  $ u_{\rm {rad}} \propto P_{\rm {rad}} \propto \rho^{4/3} \propto r^{-8/3}$, where we used the result of mass flow conservation. This allows us to connect the advective luminosity at any radius, $ L(r) = 4\pi r^2 u_{\rm rad}(r)v_{\rm w}$, to the initial luminosity $L(r_{\rm diss})$: $L(r)/L(r_{\rm diss}) \propto(r^2/r^2_{\rm diss}) \left({r}/{r_{\rm diss}}\right)^{-8/3}= \left({r}/{r_{\rm diss}}\right)^{-2/3}$, in agreement with the behaviour observed in the last panel of Fig. \ref{fig:den prof} . 
Since, $L(r_{\rm diss})/L_{\rm Edd} = (r_{\rm g}/r_{\rm diss})\left(\dot{M}_{\rm w}/\dot{M}_{\rm Edd}\right),$ the kinetic luminosity evaluated at the trapping radius becomes
\begin{gather}
\begin{split}
    L(r_{\rm tr}) = &L_{\rm Edd} \left(\frac{r_{\rm g}}{r_{\rm diss}} \frac{\dot{M}_{\rm w}}{\dot{M}_{\rm Edd}}\right)^{1/3}. 
    \label{eq:Ltr_Ledd}
\end{split}
\end{gather}
Using the characteristic values of our simulation at $t_{\rm p}$ \textendash ~ $r_{\rm diss}\approx r_{\rm p}$ (see Sec. \ref{sec:compar_sim}), $r_{\rm g}/r_{\rm p}\approx (600)^{-1}$, and $\dot{M}_{\rm w}/\dot{M}_{\rm Edd}\approx \text{several}\times10^{5}$ \textendash~ we obtain $L(r_{\rm tr}) \approx 5L_{\rm Edd}$. 
 At the trapping radius, radiation decouples from the bulk motion of the gas, and propagates outward by diffusion until it reaches the photosphere, beyond which it escapes the outflow with a luminosity  $L(r_{\rm ph})\lesssim L(r_{\rm tr})$. This behaviour is consistent with what is seen in the light curve in Fig. \ref{fig:fld}.  
 To extrapolate our scaling relations to arbitrary stellar and BH properties, rewrite Eq. \eqref{eq:Ltr_Ledd} as:
 \begin{equation}
 \label{eq: Ltr_Ledd_extended}
 \begin{split}
     \frac{L(r_{\rm tr})}{L_{\rm Edd}} \approx 6\beta^{1/3}\left(\frac{\zeta}{0.04}\right)^{1/3}&\left(\frac{\kappa}{\kappa_{\rm p}}\right)^{1/3}
     \left(\frac{M_{\rm BH}}{10^4M_\odot} \right)^{-5/18}\\
     &\left(\frac{M_\star}{M_\odot}\right)^{7/9}\left(\frac{r_\star}{R_\odot}\right)^{-5/6}\left(\frac{t}{t_{\rm fb}}\right)^{-5/9},
\end{split}
 \end{equation} 
 where $\kappa_{\rm p}=\kappa(t_{\rm p})\approx 1.44$ is the opacity at the peak of the light curve (see Sec.\ref{sec:rad}) and $\zeta$ is the wind efficiency such that $\dot{M}_{\rm w}=\zeta|\dot{M}_{\rm fb}|$ (which is $\approx 0.04$ at the peak of the light curve, see Fig.\ref{fig:MdotW}), and we assumed the wind to be launched at $r_{\rm diss} = r_{\rm p}$\footnote{We note that at $t=t_{\rm p}\approx 1.5t_{\rm fb}$ we recover $L(r_{\rm tr})\approx5L_{\rm Edd})$ estimated before}. 
 From Eq.\eqref{eq: Ltr_Ledd_extended} is clear that the appearance of an Eddington-limited luminosity occurs for certain parameter choices (and under the condition of super-Eddington dissipation rate) \textendash such as those in our simulation \textendash but is not expected to hold in other regimes (see Sec.\ref{sec:compar_sim}).
 A more comprehensive modelling is deferred to future work, where we will incorporate the evolution of the photospheric geometry and account for the diffusive layer between the trapping radius and the photosphere, where the light is actually emitted. 
 We note already that the radial velocity profile in Fig. \ref{fig:den prof} appears to decrease with radius, whereas one would expect it to undergo a brief radial acceleration and settle slowly to a terminal radial velocity $v_{\rm r} \approx \sqrt{2GM/r_{\rm diss}}\approx2\times 10^4$km/s, as most of the radiation energy has been converted into kinetic energy for the gas. The behaviour of the plotted velocity profile is an artefact of both the mass-weighted averaging and the way \texttt{RICH} handles mass, producing larger (and therefore more massive) cells at greater distances from the region of interest, namely the stream and the pericenter. While mass-weighting is appropriate for highlighting the behaviour of most of the gas, in the case of the radial velocity it disproportionately reflects the slower, nearly bound material in the accretion flow that is continually carried to larger radii by the wind.

\section{Resolution tests}
\label{sec:res_test}
\begin{figure*}
    \centering
    \includegraphics[width=0.8\linewidth]{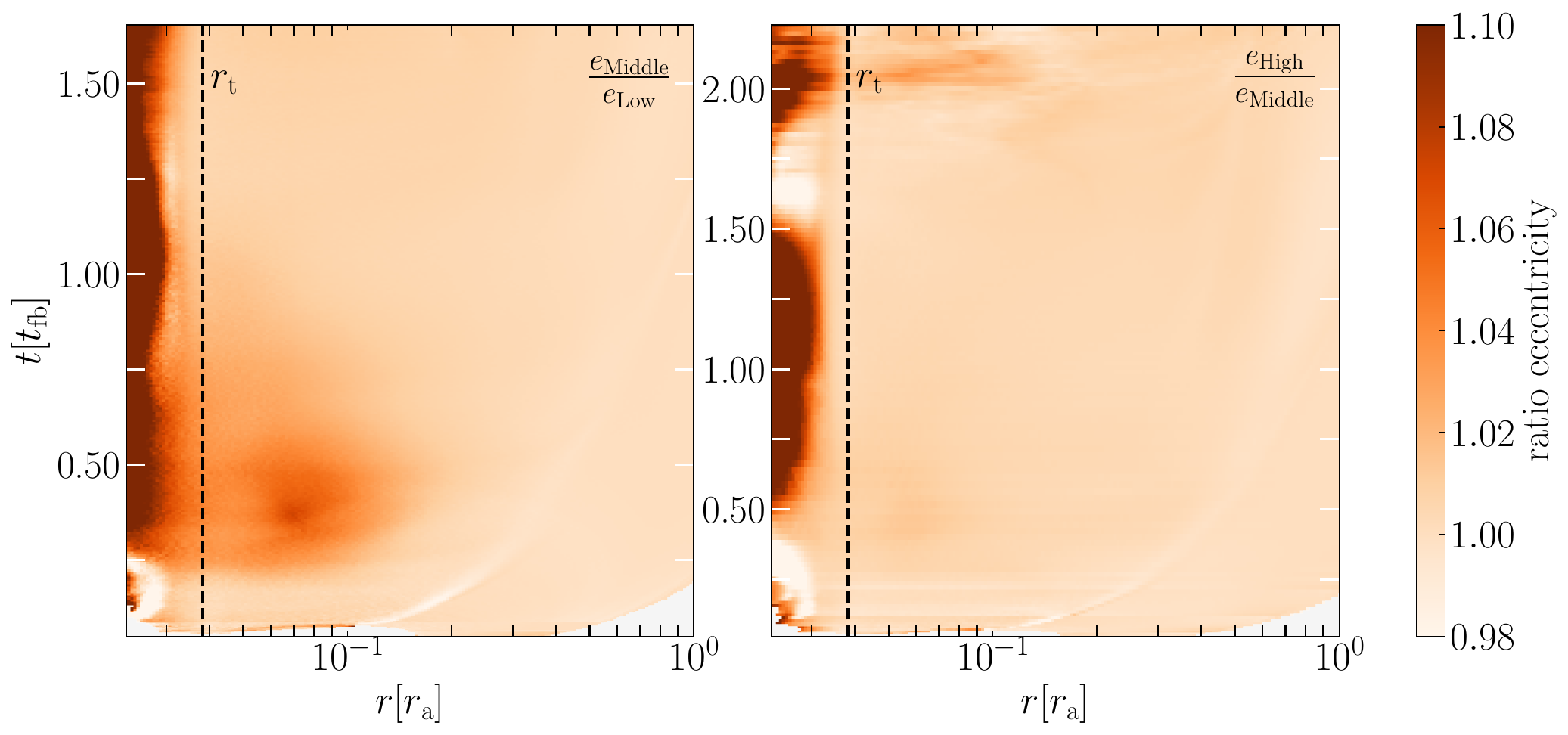}
    \caption{Space-time evolution of the ratio between gas eccentricity computed in different resolutions. \emph{Left panel}: $e_{\rm Middle}/e_{\rm Low}$. \emph{Right panel}: $e_{\rm High}/e_{\rm Middle}$. The ratio is always $\leq10\%$ outside the tidal radius (vertical  dashed line) and up to $10-20\%$ at smaller radii, indicating good convergence of bulk orbital properties.}
    \label{fig:eccrel}
\end{figure*}
In this subsection, we perform tests to evaluate the robustness of our results with respect to resolution. Although the pericenter region may be locally underresolved \citep{BonnerotLu22}, it is not clear {\it a priori} to what degree this impacts global convergence; assessing this impact is our goal here.
To quantify the discrepancies between two resolutions in computing a global quantity 
$Q$, we defined the relative error as $\mathcal{R}\equiv \max(Q_\text{res1}, Q_\text{res2})/\min(Q_\text{res1}, Q_\text{res2})$.A value of $\mathcal{R}$ approaching unity indicates that $Q$ is insensitive to resolution.
As shown in the right panel of Fig. \ref{fig:fld_R_conv}, the median photospheric radii derived for different resolutions agree to within a factor of two. A similar trend is observed in the energy dissipation rates and the radiated energy in the light curves, where $\mathcal{R} \leq 2$ after $1~t_{\rm fb}$. 
At all resolutions, we see that after a brief, mildly super-Eddington peak, the light curve settles into a luminosity consistent with $L_{\rm Edd}$. The energy dissipation rate also stabilizes at a roughly constant value, exceeding the emitted luminosity by $\approx 1.5$ orders of magnitude. The emission emerges from a photosphere that expands in time, reaching $r_{\rm ph}\approx10^{14}$cm after $2 ~t_{\rm fb}$.  Unlike dissipation rates and luminosities, the growing size of $r_{\rm ph}$ does not show signs of saturating at late times.
In line with some previous work \citep{BonnerotLu22, Kubli25}, increasing the simulation resolution decreases the dissipation rate at the nozzle at any fixed moment in time, suggesting a numerical origin for a subset of the dissipation rate.   However, the late times convergence to a uniform value of the total dissipation rate in the three runs (Fig. \ref{fig:fld_R_conv}) suggests that the eventual properties of peak luminosity and peak dissipation rate are globally converged between resolution choices (see also Appendix \ref{app: res tests}).
The discrepancies between resolutions decrease with time, suggesting that the behaviour of the \emph{High res} will closely follow that of the \emph{Middle res}. Since the latter is less computationally demanding and has therefore evolved for a longer duration, it can be considered representative of the long-term evolution of the system.

Figure~\ref{fig:eccrel} illustrates the relative errors in orbital eccentricity between different resolutions: the \emph{High res} model circularizes more slowly than the \emph{Middle res} case, which in turn circularizes more slowly than the \emph{Low res} one. Within the tidal radius, eccentricity errors are on the order of $10 - 20\%$, but they decrease to only a few percent at larger radii. Our qualitative conclusion that circularization proceeds slowly for IMBH TDEs appears robust across resolutions.

\begin{figure}
    \centering
    \includegraphics[width=\linewidth]{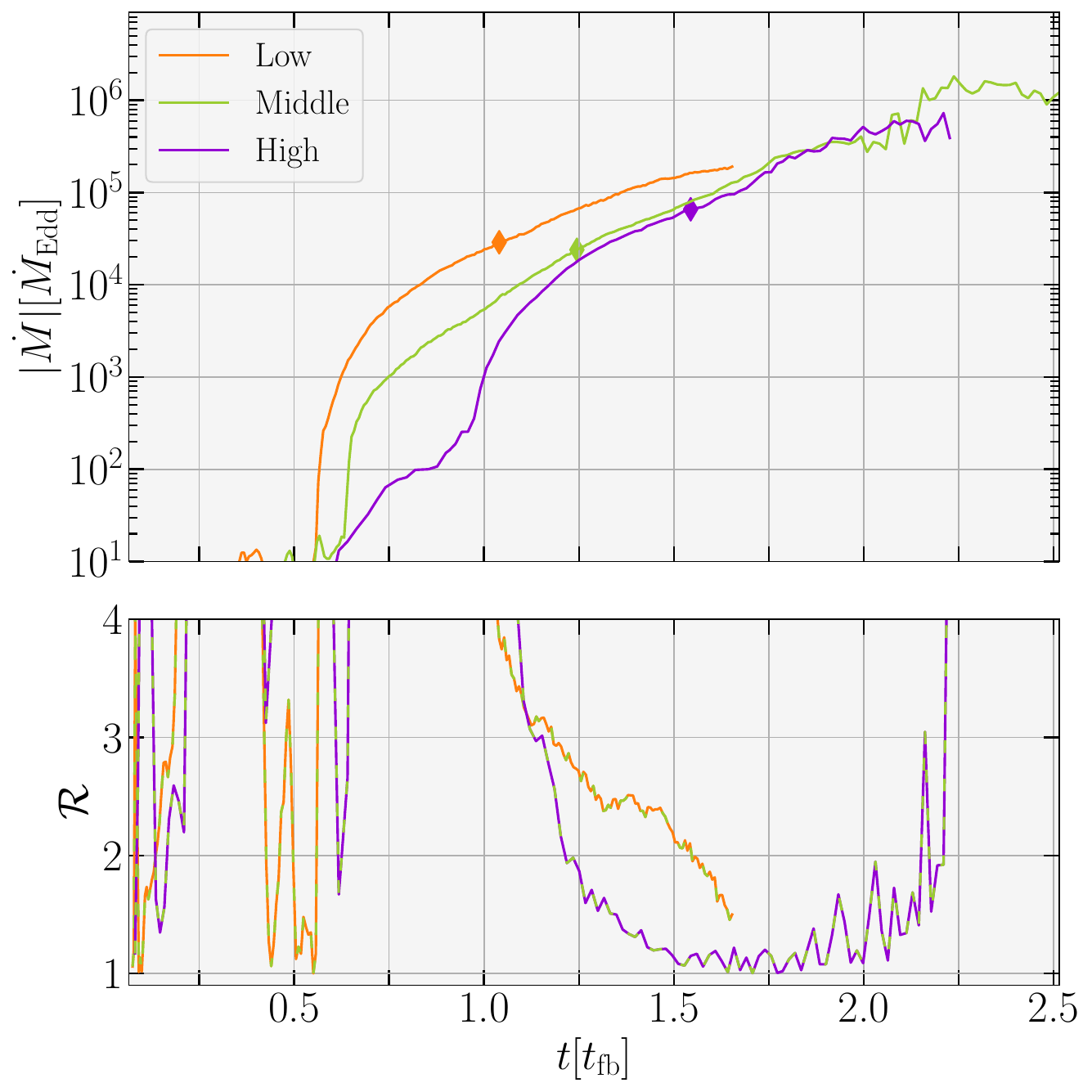}
    \caption{Resolution test for the wind analysis. \emph{Upper panel}: Time evolution of the wind mass loss rate $\dot{M}_{\rm w}$, measured at $r= 0.5 a_{\rm mb}$ in three different resolutions. Diamonds on the lines show the time $t_{\rm p}$ of the light curve peak for each resolution. \emph{Lower panel}: $\mathcal{R}$ for $\dot{M}_{\rm w}$ among two pairs of resolutions. Colours follow the scheme of Fig. \ref{fig:fld_R_conv}.}
    \label{fig:Mw_conv}
\end{figure}

We investigate convergence of wind properties in Fig.~\ref{fig:Mw_conv}.  In line with Fig. \ref{fig:fld_R_conv}, mass ejection in the wind begins earliest for the {\it Low res} case and latest for the {\it High res} case, likely due to the more rapid rise in dissipation rates at lower resolutions.  As with the light curve and dissipation rate, the total mass launching rate $\dot{M}_{\rm w}$ ultimately converges to similar values at late times (beyond 
$1.5t_{\rm fb}$).

In general, the differences between \emph{Middle} and \emph{High} resolutions are smaller than those between \emph{Low} and \emph{Middle}. Nevertheless, the global metrics of convergence we have examined here are generally reaching acceptable results at late times, despite local under-resolution of stream compression at pericenter (Martire et al. {\it in prep}). Our qualitative conclusions are  consistent across all three resolutions, with the primary (qualitative) difference being a temporal delay in saturation of global dissipation, light curve maximum, and the onset of mass loss in higher resolution runs, likely as a consequence of reduced spurious (early-time) dissipation.

\section{Discussion}
\label{sec: disc}
\begin{figure*}
    \centering \includegraphics[width=\linewidth]{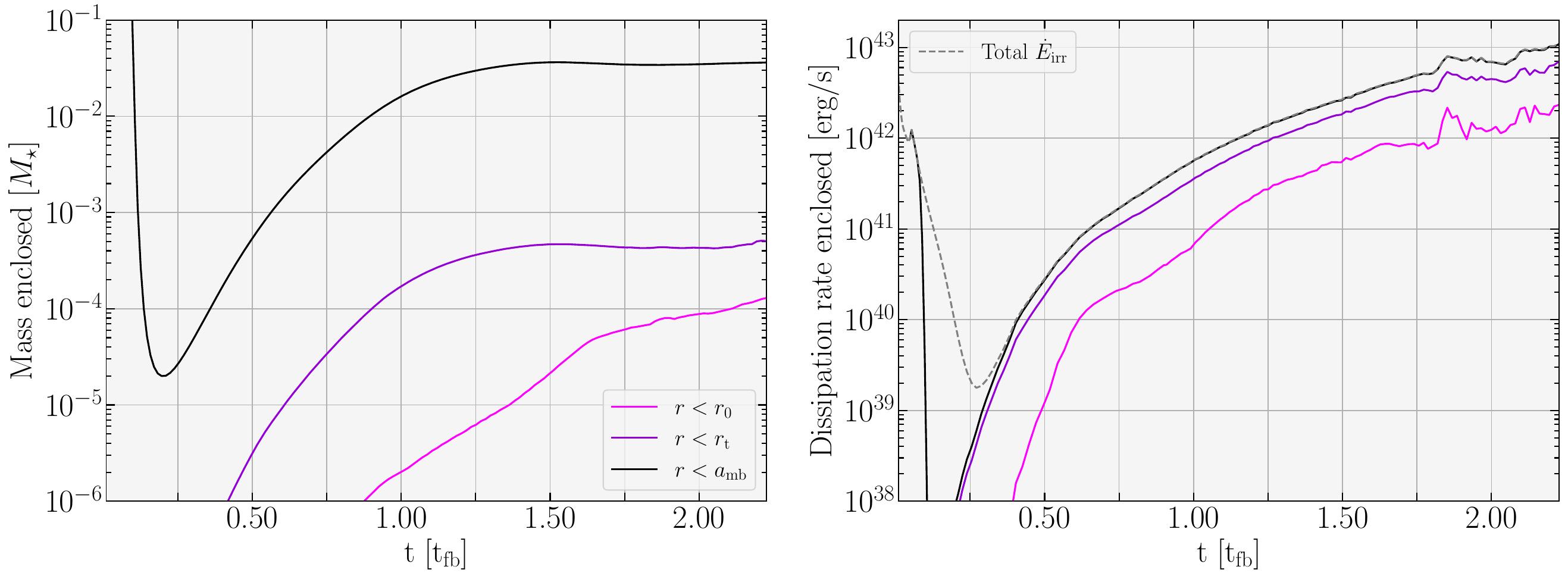}
    \caption{Material enclosed inside $r_0$ (pink line), $r_{\rm t}$ (purple line) and $a_{\rm mb}$ (black line), plotted as a function of time. \emph{Left panel}: Time evolution of the total enclosed mass. \emph{Right panel}: Time evolution of the dissipation energy rate by the enclosed material. For the entire evolution of the simulation, the dissipation energy rate for the material inside $r_0$ (pink line) is about one order of magnitude lower than the total value of the simulation (dashed gray line).}
    \label{fig:Mass encl}
\end{figure*}
The origins of early time emission in TDEs are famously uncertain \citep{Roth20}. In this paper, we have investigated the evolution of debris and the radiative emission of the disruption of a typical ($M_\star = 0.5 M_\odot$) star by an IMBH ($M_{\rm BH}=10^4 M_\odot$). Although smaller than the MBHs associated with most optically observed TDEs \citep{Hammerstein23}, our simulated system is similar to multiple X-ray-selected TDE candidates with inferred masses $\sim 10^4 M_\odot$ \citep{Lin18, Wen21, Sazonov21, Jin25, Grotova25}, whose optical emission is still poorly explored theoretically and, to a lesser extent, observationally. 
While accretion-driven evolution of the emission is likely to dominate for sufficiently late phases of a TDE (e.g., \citealt{vanVelzen19}), 
our simulations indicate, as in \citet{SS24}, that the rising phase emission is more likely to be powered by shocks.  

A major source of uncertainty in TDE modelling is the difficulty of a self-consistent simulation: only a handful of end-to-end global simulations exist in the literature for the MBH masses and stellar orbits likely to occur in observed TDEs \citep{Andal22, Ryu23, SS24, Price24}, and all of these (including the simulations presented here) likely under-resolve the nozzle shocks that initiate the process of dissipation in returning debris streams \citep{BonnerotLu22, Hu25, Kubli25}.  It is for this reason that we have performed a detailed convergence study, to identify which of our qualitative conclusions are robust to choice of resolution and which are more sensitive to the locally under-resolved nozzle shock at pericenter.  

Across all three resolutions we have simulated, we see that the first $\sim 1-2$ fallback times following an IMBH TDE are characterized by (i) very inefficient circularization, (ii) a shock-powered light curve that rises quickly and peaks near the Eddington limit, and (iii) an expanding photosphere that is embedded within an optically thick wind that advects photons generated at smaller radii. In this section, we will elaborate on caveats to our simulations, compare our results with previous hydrodynamical simulations of TDEs, and discuss our primary observational predictions.

\subsection{Validity of our assumptions}
\label{sec: grav}
Our approach has several limitations that, while necessary for computational efficiency, may introduce some inaccuracies.
First, the gravitational potential is approximated with the Paczyński–Wiita potential, 
and second, we softened it at small radii ($r<r_0$) as described in Eq.\eqref{eq:F PW}. The first of these
approximations does not significantly affect our results, since for the chosen TDE parameters $r_{\rm p}$ is 
larger than $r_{\rm g}$ by more than two orders of magnitude, and the importance of relativistic precession is minimal. The second approximation could, however, become relevant if a significant amount of mass were to accumulate at small radii and generate substantial luminosity.

Figure \ref{fig:Mass encl} shows the amount of mass enclosed within different radial cuts. Only a small amount of matter is ever located at $r\leq r_0$, reaching about $0.01\%$ of the initial stellar mass by the end of the simulation ($t \approx2.2t_{\rm fb}$). Assuming that all the matter inside $r_0$ is accreted with a radiative efficiency of $\eta= 0.05$, we can perform a {\it post hoc} check on our neglect of accretion power, by estimating its time-dependent accretion luminosity as $L_{\rm acc}=\eta\dot{M}_{\rm acc}c^2$ (with $\dot{M}_{\rm acc}$ being the net rate of mass entry into the $r\leq r_0$ region). After the peak of the light curve, $L_{\rm acc}\approx4\cdot10^{43}$ erg/s, which is higher than the shock dissipation rate in the simulation by roughly an order of magnitude. This super-Eddington value should, however, be regarded as an upper limit for accretion luminosity for several reasons.  Most importantly, the viscous time at the characteristic circularisation radius of $2r_{\rm t}$ is $t_\text{visc}\sim \Omega^{-1}\alpha^{-1}(H/R)^{-2}$, where $\alpha < 1$ is the Shakura-Sunyaev viscosity parameter and the inner disc aspect ratio is $H/R<1$.  Taking typical values from recent super-Eddington disc simulations ($\alpha \approx 0.03$, $H/R \approx 0.3$; \citealt{Jiang19}), we find that 
$t_{\rm visc} \approx 8.8$ days, or 3.4 $t_{\rm fb}$, implying that over the course of the simulation, matter accumulating near the circularisation radius will likely not have time to flow inwards\footnote{This estimate for the viscous time does, however, suggest that our neglect of the inner accretion flow would break down if we extended our simulations significantly beyond their current end-point.}.  Even if $t_{\rm visc}$ is substantially smaller than the simulation runtime (due to, e.g., an $\alpha$ value much larger than what is measured in \citealt{Jiang19}), the resulting mass inflow rate would be super-Eddington by over an order of magnitude with the assumed $\eta$, suggesting that the true $\eta$ (and thus the true accretion luminosity) will be reduced substantially by either (i) energy advection into the horizon or (ii) mass loss in a second, fast, outflow component.

Within the context of our simulation, the measured dissipation from inside $r_0$  is almost one order of magnitude lower than the total value (dashed gray line), suggesting that the simulated material at these small radii is unlikely to qualitatively change our results.
Nevertheless, future work should implement a more realistic treatment of the gravitational potential and viscosity near the black hole to achieve more accurate quantitative conclusions, particularly at late times when neglect of accretion power will eventually break down.\\

Many previous TDE simulations have often assumed an adiabatic equation of state and worked in the limit of pure hydrodynamics, neglecting both the stream’s thermal properties and the dynamic interplay between radiation pressure and radiative cooling.
The equation of state can significantly influence the evolution of the system, especially in the circularization process \citep{Hayasaki16, Bonnerot16, Jiang16, SS24}. In this work, we employ a radiative transfer algorithm to self-consistently resolve hydrodynamics and radiation together. However, our approach to radiation transport is simplified, employing a gray and diffusive approximation. This approach probably captures the bulk dynamics of debris evolution, but leads to quite approximate spectral predictions \citep{SS24}.  We are planning to alleviate this issue in future works with multi-group flux limited diffusion (Giron et al. {\it in prep}).\\

Finally, our simulations do not include magnetohydrodynamics (MHD). Although magnetic fields may play an important role in the long-term evolution of accretion discs, their influence during the early stages of TDEs is expected to be minor, as the debris streams are typically weakly magnetized \citep{Bonnerot17, Pacuraru25}. Moreover, the small number of existing MHD simulations of TDE circularization find that Reynolds stresses generally dominate over Maxwell stresses \citep{Sadowski16, Curd25, Abolmasov25}.

\subsection{Comparison to past simulations}
\label{sec:compar_sim}
The first of our primary results (slow circularization) is consistent with past TDE simulations involving even smaller IMBHs \citep{Guillochon14, Shio15}, but differs from most simulations of supermassive black hole (SMBH, i.e. $M_{\rm BH}\in[10^6, 10^8]~M_\odot$) TDEs, where circularization is seen to proceed with relative efficiency, either due to stream-stream self-intersections \citep{BonnerotLu20, BonnerotLu21}, stream-disc shocks \citep{SS24}, or a combination of both shock complexes \citep{Andal22}. It should be noted, however, that other SMBH TDE simulations find a significantly slower pace of circularization \citep{Ryu23, Price24}.  The position of our simulations in this broader landscape of circularization results seems clear: the tidal debris we simulate is unable to efficiently dissipate orbital energy at either self-intersection shocks or at smaller radii in stream-disc interactions.  Relativistic apsidal precession is weak to begin with due to the non-relativistic stellar pericenter, and dissipation efficiency at the self-intersection point near apocenter may be further weakened by asymmetries between the outgoing and ingoing streams, as in \citet{SS24}.  In contrast to the SMBH simulation of \citet{SS24}, the post-nozzle material does not seed a sufficiently dense elliptical disc capable of driving dissipation in stream-disc shocks.  The only remaining dissipation mechanism for the IMBH TDE is the nozzle shock, which (while capable of producing Eddington levels of luminosity) is far too weak \citep{Hu25} to circularize debris on a timeframe of $\sim t_{\rm fb}$.

The second of our results (Eddington-limited bolometric luminosity) is likewise consistent with previous simulations of SMBH TDEs \citep{BonnerotLu21, Huang24, SS24, Price24, Curd21}.  These SMBH simulations find peak luminosities that are Eddington-limited to within a factor of a few, despite peak fallback rates that can be super-Eddington\footnote{This characterization of the Eddington ratio of the mass fallback rate assumes a thin disc radiative efficiency of $\approx 0.1$.} by an order of magnitude.  While the Eddington-limited peak luminosities of SMBH TDEs are perhaps not so surprising, this is a more unexpected result for our IMBH simulations: naive mass fallback rates in an IMBH TDE are roughly $\sim 10^5 \times$ the Eddington limit\footnote{Again, assuming a thin disc radiative efficiency.}, and shock power is not intrinsically Eddington limited in the manner that accretion power is.  The low dissipation efficiency of the nozzle shock resolves this puzzle partially, but not fully: peak dissipation rates in our simulations exceed emitted luminosities by a factor $\approx 30$.

The origins of the Eddington limit in our simulations seem tied to the nature of our photosphere.  Radiation energy is emitted through a radiation-driven wind that expands (and also becomes more spherical) in time. The radial density profile scale as $r^{-2}$ as \citet{Jiang16}, and, as our analytic model in Sec.\ref{sec: wind} shows, this type of outflowing photosphere will generally produce near-Eddington luminosities when the system parameters fall within a specific regime, as the one of our simulation. 
If we extrapolate Eq. \eqref{eq: Ltr_Ledd_extended} to a range of other MBH masses not simulated here, we would predict that most main sequence TDEs should ultimately saturate at near-Eddington peak luminosities.  For example, Eq. \eqref{eq: Ltr_Ledd_extended} predicts $L(r_{\rm tr}) = \{7.4, 3.9, 2.1, 1.1 \} L_{\rm Edd}$ for $M_{\rm BH} = \{10^3, 10^4, 10^5, 10^6\} M_\odot$ at $t=t_{\rm fb}$ (assuming disruption of the star simulated here, with $\kappa=\kappa_{\rm T}$ and $\zeta=0.04$).  As noted in Sec.\ref{sec: wind}, this convergence to near-Eddington values is both surprising (as it holds even for IMBH disruptions where shock dissipation rates may be orders of magnitude super-Eddington) and something of a coincidence: we predict $L(r_{\rm tr})/ L_{\rm Edd} \propto M_{\rm BH}^{-5/18}$, and therefore that the BH dependence of the advected luminosity is weak.  This prediction of our simple wind model is in agreement with the results of \citet{SS24}, which found shock-powered and near-Eddington emission from a $10^6 M_\odot$ disruption.  However, the ratio $L(r_{\rm tr})/L_{\rm Edd}$ has a significant dependence on stellar parameters.  If we had considered TDEs from radically different stars, for example the disruption of a white dwarf or of a $100 M_\odot$ main sequence star, then $L(r_{\rm tr})$ would be one or more orders of magnitude higher.  Eq. \eqref{eq:Ltr_Ledd} would not apply if initial mass fallback rates yield sub-Eddington dissipation rates, as would be expected for partial disruptions or very large ($M_{\rm BH}\gtrsim 10^{6.5} M_\odot$) SMBH disruptions of main sequence stars.\\


\begin{figure}
    \centering
    \includegraphics[width=\linewidth]{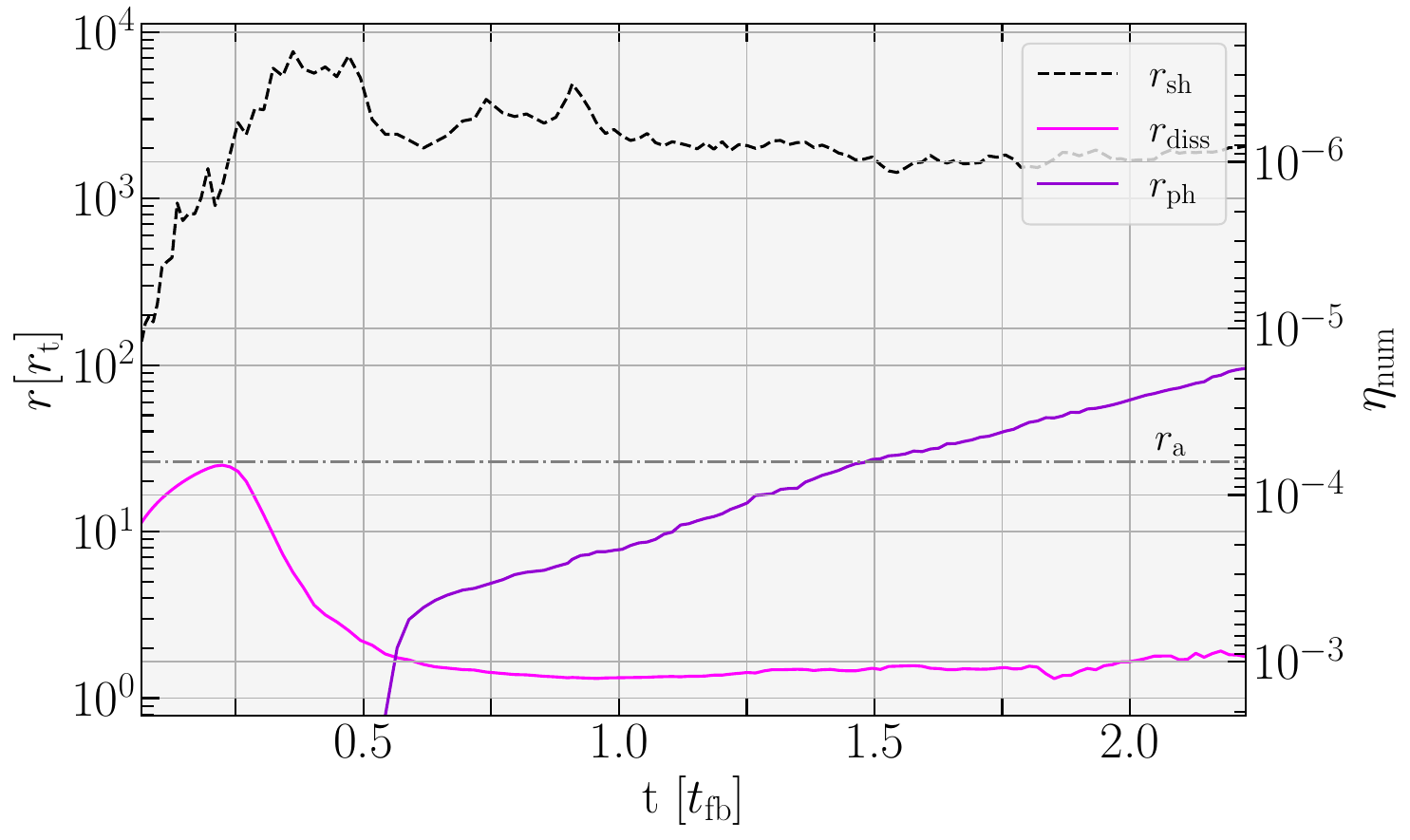}
    \caption{Radii of observable interest for the \emph{High} resolution. The solid violet line is the photospheric radius, computed as a median value over different lines of sight. The dissipation radius (magenta line) indicates the physical site where shock dissipation is maximized; conversely, the ``shock radius'' (dashed black line) shows the dissipation site that would be inferred under the assumption that luminosity is produced by efficient stream self-intersections at $r=r_{\rm sh}$. We see that dissipation is always dominated by the nozzle shock, near pericenter, and well inside the photosphere.  The right-hand-side $y$-axis is defined as $\eta_{\rm num}=r_{\rm g}/r$.  This converts $r_{\rm sh}$ and $r_{\rm diss}$ into dimensionless ``radiative efficiencies'': for $r_{\rm sh}$, this represents the true efficiency with which returning rest mass energy is converted into radiation ($\eta_{\rm num}= L_{\rm FLD}/(\dot{M}_{\rm fb}c^2)$). For $r_{\rm diss}$, $\eta_{\rm num}$ is the (far higher) radiative efficiency that would be achieved if the nozzle shock thermalized and radiated all the kinetic energy of returning debris streams.  Interpreting the emergent luminosity as originating from efficient ($\eta \sim r_{\rm g} / r_{\rm sh}$) apocentric shocks yields unphysically large shock radii (the implied $r_{\rm sh}$ would be $\gg r_{\rm a}$).}
    \label{fig:Rshock_eta}
\end{figure}
We summarize our three resolution-insensitive results and their interplay by plotting characteristic radii in Fig.\ref{fig:Rshock_eta}. We tracked energy dissipation defining the dissipation radius $r_\text{diss}$ as
\begin{equation} 
\label{Rdiss}
    r_\text{diss}=\frac{\sum_i \dot{E}_{{\rm irr},i}r_i}{\sum \dot{E}_{{\rm irr},i}},
\end{equation}
where the sum is done over all the simulation cells and $\dot{E}_{\rm irr}$ is computed as in Eq.\eqref{eq: uirr}. Fig.\ref{fig:Rshock_eta} shows that at early times energy is dissipated in the bulk of the disrupted star.
After approximately $0.75t_{\rm fb}$, it localizes around the pericenter as the nozzle shock forms, converting kinetic energy of the infalling material into thermal and radiation energy and launching the wind described in Sec.\ref{sec: wind}. 
The nozzle shock can impart a large width difference to the stream \citep{BonnerotStone21}, influencing the subsequent formation and evolution of additional shocks and the accretion disc. In agreement with \citet{Shio15}, there is no evidence of circularization yet and no stream-disc shock or secondary shock can happen, contrary to what has been seen in other simulations \citep{SS24, Piran15}. We leave further exploration to future works. 
If the kinetic energy of matter in an accretion flow is entirely converted into radiation through shocks \citep{Piran15, Ryu20}, the corresponding effective radius $r_\text{sh}$ of the surface that, with a mass inflow rate $\dot{M}_{\rm fb}$, would generate the luminosity $L_{\rm FLD}$ from Eq.\eqref{eq:L fld} would be:
\begin{equation}
    \label{eq: LmaxRsh}
    r_\text{sh} = \frac{GM_\text{BH}\dot{M}_\text{fb}}{L_{\rm FLD}}, 
\end{equation}
which (with our measured value of $L_{\rm FLD}$) is many orders of magnitude\footnote{The result is independent on resolution.} larger than $r_\text{diss}$, and even orders of magnitude greater than $r_{\rm a}$. 
This mismatch in radii reflects the very low efficiency of dissipation $\eta_{\rm sh}= L_{\rm FLD}/(\dot{M}_{\rm fb}c^2)$ seen in our simulations: actual dissipation rates are orders of magnitude lower than $GM_{\rm BH} \dot{M}_{\rm fb}/r_{\rm p}$, even though the physical locus of dissipation sits at radii $r_{\rm diss} \sim r_{\rm p}$.\\

Simulations of IMBH TDEs have long been understood to be more computationally tractable than SMBH TDEs on grounds of dynamic range \citep{Ramirez-Ruiz09, Guillochon14, Shio15}.  Our results indicate that these encounters may be more computationally tractable along other dimensions as well.  While the nozzle shock in our simulations is under-resolved, global results are generally insensitive to numerical resolution.  The dynamics of the nozzle are likely to matter more in SMBH TDEs, where the ``exit angle'' with which stream material leaves the nozzle can determine \citep{BonnerotStone21} the initial mode of circularization: via self-intersections if spreading is minimized \citep{BonnerotLu20, Andal22}, or via stream-disc interactions if spreading is large enough \citep{SS24}.  In IMBH TDEs, at least at early times, the lack of any appreciable circularization reduces the importance of nozzle resolution.  Likewise, the unimportance of stream-disc interactions may make IMBH TDEs more easily simulated by mass injection schemes \citep{BonnerotLu20, BonnerotLu21, Huang24}, which can be more computationally efficient but sacrifice the ability to fully capture stream-disc interactions.  If these are not relevant for IMBH TDEs, then stream injection simulations will be much more robust.

\subsection{Observational implications}
Presently, the rate of TDE discovery is dominated by wide-field optical time domain surveys \citep{SjoertUV}, most notably ZTF \citep{Sjoert21, Hammerstein23}.  These surveys have not found strong candidate TDEs from MBHs far below $\sim 10^6 M_\odot$.  Within the ZTF TDE sample, the smallest MBH mass inferred from galaxy scaling relations is $\sim 10^{5.5} M_\odot$ \citep{Yao+23}.  Although the applicability of scaling relations is questionable in the IMBH range, this finding is corroborated by the lack of ZTF TDEs with very fast rise times; the shortest observed rise is $\approx 6.4$ days, and this is an outlier \citep{Yao+23}.  The situation is different for the population of X-ray selected TDEs, where multiple IMBH candidates exist \citep{Maksym14, Lin18, Sazonov21, Wen21, Grotova25}.    

The estimated median temperature at the photosphere (Fig.\ref{fig:fld}) will produce optical to UV emission broadly consistent with observations.
A quasi-spherical emission surface is supported by polarimetry observations of some TDEs \citep{Leloudas22, Patra22, Wichern25}. 
The asymmetry of the photosphere at early times, however, suggests that there might be different spectral energy distributions when varying lines of sight, reminiscent of the viewing angle model \citep{Dai18}, with the important difference that at this time in our simulations, no accretion disc has formed.

Given the peak luminosity of our simulated light curve, $L(t_{\rm p})\approx 8\cdot10^{41}$ erg~s$^{-1}$, we can approximately estimate the maximum redshift $z_{\rm max}$ at which similar TDEs can be observed, particularly by upcoming surveys. We compute the observed flux
\begin{equation}
    F_{\rm obs}(z) = \frac{L(t_{\rm p})}{4\pi D^2_{\rm L}(z)},
\end{equation}
where $D_{\rm L}$ is the luminosity distance assuming $H_0=70$km~s$^{-1}$~Mpc$^{-1}$, $\Omega_{\rm M}=0.3, ~\Omega_\Lambda=0.7$.  Assuming black body emission at the median radiation temperature of the photosphere\footnote{$T=T(t_{\rm p})\approx 2.5\cdot10^4$K, as can be seen in the left panel in Fig.\ref{fig:fld}.}, the in-band flux is given by $F_\nu(z) = F_{\rm obs} B(\nu_{\rm c}, T)\pi/(\sigma_{\rm SB}T^4)$, where $\nu_{\rm c}$ is the central frequency of the band.
The apparent AB magnitude is given by $m_{\rm AB}(z)= -2.5\log_{10}(F_\nu/(3631 \text{Jy}))$.
Telescopes such as the Vera Rubin Observatory (limiting magnitude $m_{\rm AB, r} =24.7$ mag for single exposure; \citealt{Bianco22}) 
will be able to observe events like our simulated TDE up to redshift $z\approx0.1$; the upcoming {\it ULTRASAT}~ UV sky survey satellite (limiting magnitude $m_{\rm AB}\approx22.4$ mag; \citealt{Shvartzvald24}) will have a smaller horizon of $z\approx0.06$, although we emphasize that our estimates of colour temperature are quite approximate. Assuming a volumetric rate of $10^{-6}\text{Mpc}^{-3}\text{yr}^{-1}$ \citep{Yao+23}, this corresponds to roughly 100 events per year detectable by LSST and about 15 per year for {\it ULTRASAT}\footnote{The true detection rate could be lower if IMBHs have much lower volumetric TDE rates than SMBHs, as has been suggested by some recent loss cone modelling \citep{Chang+25, Hannah+25}.}.

On the other hand, the non-detection of TDEs from $10^4M_\odot$ BHs by existing surveys is consistent with expectations. The ZTF survey ($m_{\rm AB, r} = 20.5$ mag, \citealt{Graham19}) can only detect such events out to $D_{\rm L}\approx65 \text{Mpc}$ ($z\approx0.015$), which coincides with the distance of the nearest optically discovered TDE in the ZTF sample (see Table 3 in \citealt{Yao+23}). Therefore, unless these events are significantly more luminous, much more common than SMBH TDEs, or cooler than we have estimated (shifting more flux into the optical band), their detection by ZTF remains unlikely.

Similar though more uncertain conclusions can be drawn for {\it SRG/eROSITA} survey. The fraction of the bolometric power emitted in soft X-rays $f_{\rm X}$ is poorly constrained by our simulation, which neglects the inner regions of any accretion flow that could have formed. If we assume $f_{\rm X}=10^{-1}$ for soft X-ray bands, then {\it eROSITA} (whose average flux limit of $F_{\rm obs}\approx 8\cdot10^{-14}$ erg~cm$^{-2}$s$^{-1}$ for $0.2-2.3\text{keV}$ photons \citep{Predehl21} implies a TDE detection flux threshold of $F_{\rm obs}\approx 3\cdot10^{-13}$ erg~cm$^{-2}$s$^{-1}$ , adopting the variability selection criteria of \citet{Grotova25a, Grotova25}) would have an X-ray horizon of only $\approx 50 ~\text{Mpc}$ ($z\approx 0.01)$, of the order of the nearest TDE detected by {\it eROSITA} \citep{Grotova25}. 
For the recently launched {\it Einstein Probe/WXT}, discovering classical IMBH TDEs in X-rays will be even more challenging since its wide-field flux limit is significantly higher ($F_{\rm obs}\approx 2.5\cdot10^{-11}$ erg~cm$^{-2}$s$^{-1}$ in the $0.5-10\text{keV}$ band \citep{EinsteinP25}, corresponding to a maximum detection distance of only $5~ \text{Mpc}$). Nonetheless, the survey’s large grasp may still enable the discovery of more exotic or unusually bright TDEs (e.g., jetted TDEs) which could provide valuable physical insight into these transients. 
More accurate detectability estimates would require detailed spectral modelling from simulations, which is beyond the scope of this paper. \\

Observations have found that radio afterglows are quite common in TDEs \citep{Alexander+25}, likely originating in outflows that create shocks and accelerate relativistic electrons with the circumnuclear medium.  
Although most TDEs only brighten in the radio at late times \citep{Cendes+24}, suggesting a delayed outflow launch, a minority of TDEs show prompt radio emission that is likely driven by the interaction between the circumnuclear medium and unbound material ejected prior to the optical peak (\citealt{Goodwin_2_23, Goodwin23}; see especially Fig. 4 of \citealt{Alexander+25}). As shown in Figures \ref{fig:MdotW} and \ref{fig:orbE_IE_relDiff}, the amount of unbound material increases over time, releasing energies on the order of $10^{49}$ erg. 
Its collision with the surrounding medium could play an important role in generating synchrotron emission at larger radii. Further investigation of the early time wind properties will be essential to fully understand the origin of radio emission in TDEs. \\

Finally, we mention that a separate class of flares, Luminous Fast Blue Optical Transients (LFBOTs), may be related to TDEs. They are characterized by extreme peak optical luminosities ($>10^{43} \text{erg/s}$, \citealt{Drout14}) and fast rise/decay in the optical light curve (few days). Their origin is still unknown. Different models have been proposed to explain them \citep{Migliori24}, such as rapidly-rotating magnetars, accreting BHs born in failed blue supergiant star explosion, and IMBH ($\sim 10^3-10^4 M_\odot$) TDEs \citep{Perley19}.
The bolometric luminosity we derive is substantially lower than the optical peaks of LFBOTs, disfavouring early time IMBH TDEs as the dominant explanation for LFBOTs. Nevertheless, since some IMBH TDE disc models can 
match late-time behaviour of LFBOTS  \citep{Inkenhaag25}, a more detailed investigation is needed, involving synthetic spectral modelling of our simulations and a better treatment of the inner accretion flow.

\section{Conclusions}
\label{sec:concl}
We have simulated the tidal disruption of a $0.5M_\odot$ polytropic ($n=1.5$) star on a parabolic orbit by a $10^4M_\odot$ IMBH, following the evolution from the event itself up to $\approx 2$ dynamical times. We evolve the stellar debris with a realistic equation of state in 3D radiation hydrodynamics.  We find the following results in our fiducial ({\it High res}) simulation:
\begin{enumerate}[(i)]
    \item Radiation is advected by an anisotropic outflow that forms near pericenter and becomes more spherical and unbound with time (see Fig. \ref{fig:fld});
    \item the photosphere continues to expand throughout the simulation, reaching radii of $r\approx r_{\rm a}\approx10^{13}$cm with a temperature of few $\times10^4$K at the peak of the light curve, and growing even further to $r\approx4r_{\rm a}\approx10^{14}$cm after 2$t_{\rm fb}$ (see Figs.~\ref{fig:denproj} and \ref{fig:fld}).
    \item The light curve peaks at roughly 2$L_{\rm Edd}$ before declining and then levelling off at $L_{\rm Edd} \approx 3 \times 10^{41}~{\rm erg~s}^{-1}$ (see Fig.~\ref{fig:fld}).
    Both the radiated luminosity and the energy dissipation rate are multiple orders of magnitude below the classical estimates for IMBH TDE luminosities given by $L=0.1\dot{M}_{\rm fb}c^2$ (which, for our parameters, would be approximately $7\times10^{46}~{\rm erg~s}^{-1}$). The result of luminosity being capped by the Eddington limit holds for most main sequence TDEs with different MBH masses (under the condition of super-Eddington dissipation rate), but not for all stellar parameters.
    \item Circularization is slow and the overwhelming majority of the material is still on eccentric orbits after the peak of the light curve (see Fig. \ref{fig:ecc}).
\end{enumerate}
We assessed the robustness of our results by running convergence tests and found that changing the resolution of the simulation does not change our qualitative conclusions.  The compression of returning streams at the pericentric nozzle may be locally under-resolved, as fully resolving the shocks here would require cell sizes $\sim10^{-3}-10^{-4} R_\odot$ \citep{BonnerotLu22}. This could affect the conclusions on the effects of the nozzle shock and the estimates of the width of the stream.  

Our approach has several limitations, which we intend to overcome in future works. First, our treatment of gravity relies on a pseudo-Newtonian prescription and incorporates an artificially softened potential at small radii. We checked the effect of these approximations in Sec.\ref{sec: grav} and concluded that they do not qualitatively impact the results. However, a more accurate gravitational model would be necessary to study highly relativistic TDEs (where $r_\text{p}\approx r_\text{g}$) or to simulate their long-term evolution. Second, our approach to radiation transport is simplified, employing a gray and diffusive approximation. 
Third, we do not include magnetic field, but the deviation from our results would be negligible due to the parameters employed for our simulation. 

Despite these caveats, in this paper we have shown that energy is mainly dissipated in the pericenter region, where a radiation-driven wind forms and expands quasi-spherically. The radiation is carried outward by this flow, producing an Eddington-limited luminosity consistent with analytical predictions. To first order approximation, next-generation surveys as LSST and {\it ULTRASAT} are expected to detect 10 to 100 IMBH TDEs per year. Future simulations with higher resolution and more sophisticated radiative modelling will yield more accurate predictions and enable a more detailed characterization of the stream and outflow evolution, better connecting fundamental TDE parameters to observables.

\section*{Acknowledgements}
The authors thank Yujie He, Vysakh Anilkumar, Pietro Baldini, Simona Pacuraru and John Ryan Westernacher-Schneider for useful discussions. EMR and PM acknowledge support from the European Research Council (ERC) grant number: 101002511/project acronym: VEGA\_P.  NCS acknowledges support from the Binational Science Foundation (grant No. 2020397) and the Israel Science Foundation (Individual Research Grant No. 2414/23). IL acknowledges support from a Rothschild Fellowship and The Gruber Foundation, as well as Simons Investigator grant 827103. 
This publication utilises computing time on the Dutch National Computer Facility, Snellius. It is part of the project ``A Library of Disruption Event Simulations: the key to Intermediate Mass Black Hole Discovery'', with file number 022.016, which is funded by the Dutch Research Council (NWO). EMR also acknowledges that this publication is part of the project ``Flares from disrupted stars unveil the origin of the giant black holes'' with file number  VI.C.232.099 of the research programme VICI, which is financed by NWO.

\section*{Data availability}
The data underlying this article will be shared on reasonable request to the corresponding author.

\bibliographystyle{mnras}
\bibliography{biblio} 


\appendix
\renewcommand{\thefigure}{\Alph{section}\arabic{figure}}
\setcounter{figure}{0}

\section{Numerical details}
\subsection{Gradient calculations}
\label{app: grad}
In this appendix we present the details for the calculations in  Eq.\eqref{eq: grad} and \eqref{eq: F results fld}.  For each observer at position $\textbf{r}$ from the BH, we need to compute the gradient of the radiation energy density $u_\text{rad}$ at the photosphere and the related quantities
\begin{equation}
\begin{split}
    \nabla_\text{r}u_\text{rad}&= \nabla u_\text{rad} \cdot \hat{\textbf{r}},\\
    |\nabla u_\text{rad}| &= \sqrt{(\nabla_\text{x} u_\text{rad})^2 + (\nabla_\text{y} u_\text{rad})^2 +(\nabla_\text{z} u_\text{rad})^2},
\end{split}
\end{equation}
where $\hat{\textbf{r}}=\textbf{r}/|\textbf{r}|$, while $\nabla_\text{x}, \nabla_\text{y}, \nabla_\text{z}$ denote the projection of the gradient on the 3D Cartesian directions, whose calculation is described in the following.
For each photospheric cell $(\bar{x},\bar{y},\bar{z})$ of size $s = V^{1/3}$ (with $V$ being the cell volume), 
\begin{equation}
\label{eq:grad_app}
\nabla_\mathrm{x} u_\text{rad} = \frac{u_\text{rad}\left(\bar{x} + s, \bar{y}, \bar{z}\right) - u_\text{rad}\left(\bar{x} - s, \bar{y}, \bar{z}\right)}{2s},
\end{equation}
where we used the twenty nearest cells to interpolate the simulation data for radiation energy density in the numerator of Eq.\eqref{eq:grad_app}.
Analogously, we computed $\nabla_\mathrm{y} u_\text{rad}, \nabla_\mathrm{z} u_\text{rad}$.\\
We mention that the flux for each observer is not the direct result of Eq.\eqref{eq: F results fld} at the photosphere, but a uniform moving mean along the six nearest neighbours.

\subsection{Dissipation rates}
\label{app: diss}
While self-consistently solving for both radiation and hydrodynamics, \texttt{RICH} tracks the total energy dissipated. For energy conservation (neglecting gravity, and viscous dissipation\footnote{Radiation is not considered here, since it is treated separately in the radiation step, where it is linked to the change in gas kinetic energy.}),
\begin{equation}
\label{eq app: E con}
    \frac{\partial u}{\partial t} + \vec{\nabla}\cdot(u \vec{v}) = - \vec{\nabla}\cdot(P\vec{v}),  
\end{equation}
where $u\equiv\rho(\varepsilon_{\rm th}+v^2/2)$ is the total energy density. The total rate of energy change is found integrating over the cell volume\footnote{Here we used the Reynolds transport theorem and the Gauss theorem.} as
\begin{equation}
\begin{split}
    \dot{E}_{\rm tot} 
    &\equiv \frac{{\rm d}}{{\rm d}t} \int_V u\, {\rm d}V 
    = \int_V \left[ \frac{\partial u}{\partial t} + \boldsymbol{\nabla} \cdot (u \boldsymbol{v}) \right] {\rm d}V \\
      & = - \int_V \boldsymbol{\nabla} \cdot (P \boldsymbol{v})\, {\rm d}V 
    = -\oint_{\partial V} P\, \boldsymbol{v} \cdot \boldsymbol{n}{\rm d}S, 
\end{split}
\end{equation}
which is approximated by \texttt{RICH} as
\begin{equation}
\label{eq app: E tot}
    \dot{E}_{\rm tot} \approx -\sum_{\text{interfaces}} A\, P_*\, \boldsymbol{v}_* \cdot \boldsymbol{n}_{\rm A},
\end{equation}
where $A$ is the area of the interface (and $\vec{n}_{\rm A}$ its normal vector) and the starred quantities are computed at the contact discontinuity through Riemann solvers\footnote{We note that Eq.\eqref{eq app: E con} is ill-defined in case of shocks and one should adapt directly the integral form and the weak solutions \citep{toro09}. However, this would not change the final approximation given in Eq.\eqref{eq app: E tot}}. 

Assuming only reversible work (and thus a smooth flow), one can write $\vec{\nabla}\cdot(P\vec{v})  =  P\vec{\nabla}\cdot\vec{v} + \vec{v}\cdot\nabla P$ and integrating \eqref{eq app: E con} \citep{toro09} gives
\begin{equation}
\begin{split}
\label{eq:adia} 
     \dot{E}_{\rm rev}&\equiv\frac{\rm d}{\rm dt} \int u \rm{d}V 
     = - \int \vec{\nabla}\cdot(P\vec{v}) \rm{d}V 
     = - \int ( P\vec{\nabla}\cdot\vec{v} + \vec{v}\cdot\nabla P)\rm{d}V \\
    &= - P_c\oint \vec{v} \cdot \rm{d}\vec{A} - \vec{v}_c\cdot \vec{n}_{\rm A}\oint P \rm{d}A  \\
     &\approx- \sum_{\text{interfaces}} A \Big(P_c \vec{v}_*+ \vec{v}_c P_*\Big)\cdot\vec{n}_{\rm A},
\end{split}
\end{equation} 
where $P_{\rm c},\vec{v}_{\rm c}$ are gas pressure and velocity at the cell centre.
Therefore, the rate of change of the dissipation energy is
\begin{equation}
     \dot{E}_{\rm irr} = \dot{E}_{\rm tot} - \dot{E}_{\rm rev} = -\sum_{\text{interfaces}} A (P_*\vec{v_*}-P_c\vec{v_*}-\vec{v}_cP_*)\cdot \vec{n}_{\rm A},
\end{equation}
as given in Eq.\eqref{eq: uirr}. 
\begin{figure*}
    \centering
    \includegraphics[width=\linewidth]{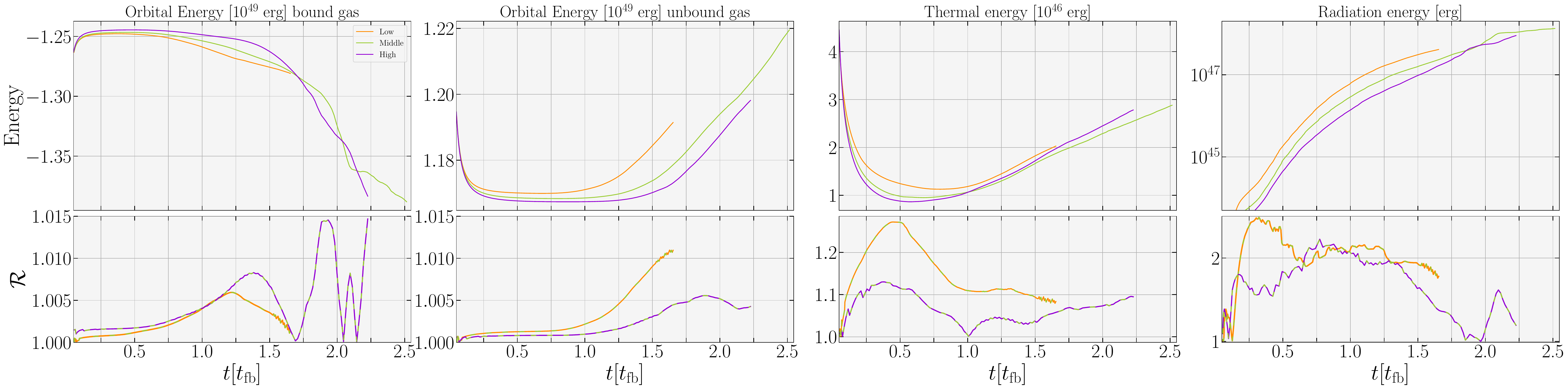}
    \caption{Resolution sensitivity test for the energy budget. \emph{Upper panels}: Time evolution of energies in different resolutions. \emph{Lower panels}: Time evolution of $\mathcal{R}$ (see Sec. \ref{sec:res_test}) for different energies. \emph{From left to right}: orbital energy for bound gas, orbital energy for unbound gas, gas thermal energy, radiation energy. Colours follow the scheme of Fig.\ref{fig:fld_R_conv}. Resolutions are highly converged in the evolution of orbital energy, and converged at the $\sim 10-15\%$ level in the evolution of gas thermal energy and radiation energy.  Convergence is weakest just after the most bound tip of the stream returns to pericenter.}
    \label{fig:orbE_IE_relDiff}
\end{figure*}
\begin{figure}
    \centering
    \includegraphics[width=\linewidth]{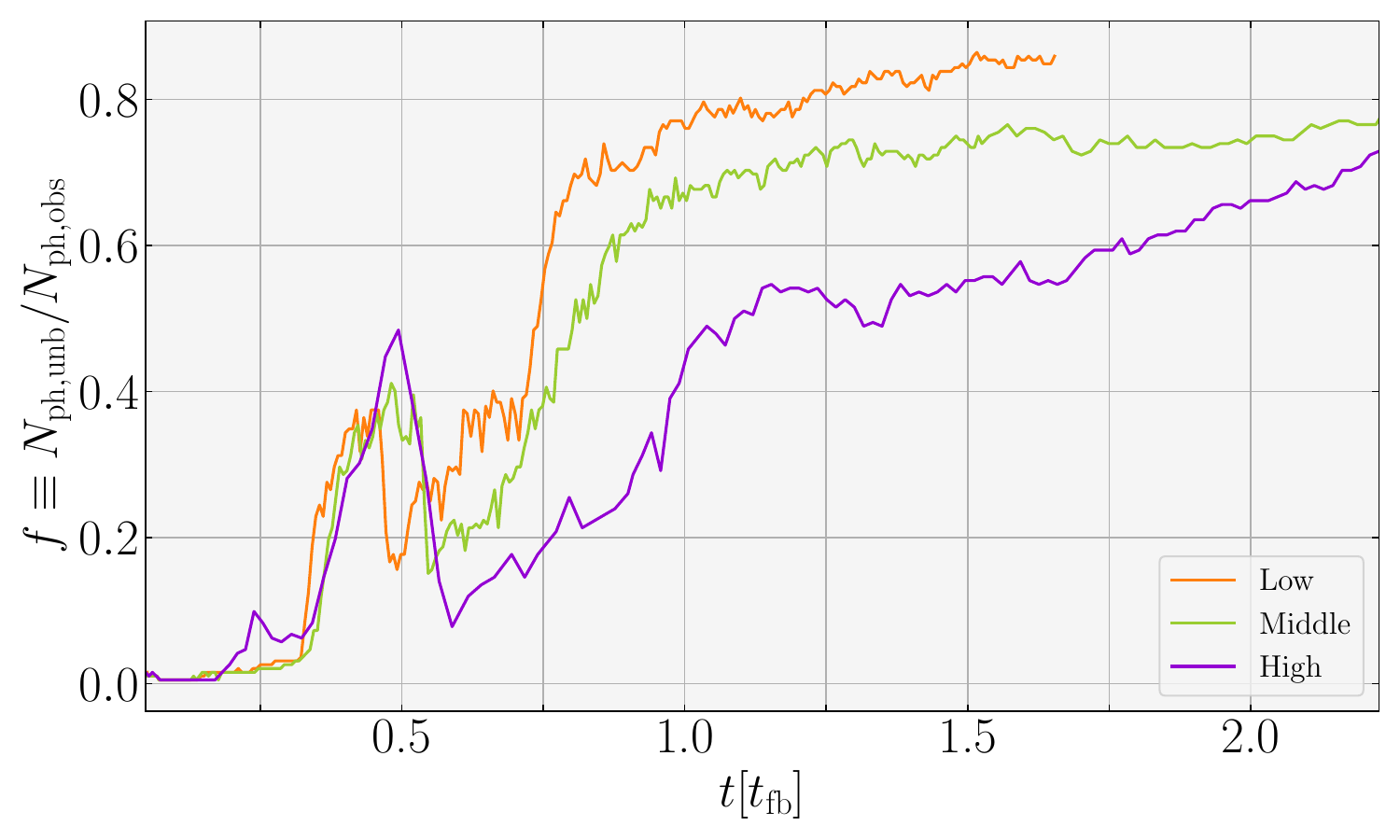}
    \caption{Fraction $f$ of unbound lines of sight at the photosphere over the total $N_{\rm obs}$ in different resolutions. Colours follow the scheme of Fig.\ref{fig:fld_R_conv}.}
    \label{fig:f_conv}
\end{figure}
\section{Mean absorption coefficients}
\label{app: opacity}
We use tabulated values for mean absorption coefficients following \citet{Krief16} within the range $[10^{-10}-10^2]\text{g/cm}^3$ for density and $[6\cdot10^3-6\cdot10^7]K$ for temperature\footnote{The tables used can be found in \url{https://gitlab.com/eladtan/RICH/-/tree/master/data/STA?ref_type=heads}.}. 
For the Rosseland mean absorption coefficient $\alpha_\text{Ross}$, we performed a bilinear extrapolation in log-space for density and temperature outside the tabulated range using the boundary point and the seventh point inward. This approach includes the following exceptions:
\begin{enumerate}[(i)]
    \item For temperatures exceeding the tabulated maximum, $\alpha_\text{Ross}$ is assumed to be independent of temperature.
    \item Compton scattering is used as a lower limit at low densities ($\rho < 10^{-10}$g/cm$^{3}$) or at high temperatures ($T > 6 \cdot 10^7$K).
\end{enumerate}
We used tabulated values for the Compton scattering coefficient $\sigma$ and extrapolated it bilinearly in log-space using the relation $\ln \sigma = \ln \rho + B \ln T$, where the coefficient $B$ is determined from the boundary point and the seventh point inward. For temperatures above the tabulated maximum, we set $B = 0$, assuming $\sigma$ to be independent of temperature.

\section{2D projections}
\label{app: denproj}
To generate the density projection in Fig.\ref{fig:denproj}, we considered a three-dimensional sample of points $\text{S}=S_x\times S_y\times S_z$ (where $S_x=[-6, -2.5]r_\text{a},\, S_y = [-4,3]r_\text{a}, \,S_z = [-2, 2]r_\text{a}$ are arrays of, respectively, $N_\text{x}= N_\text{y}=800, \,N_\text{z}=100$ linearly spaced points) and searched for the simulations cells closest to them using the \texttt{KDTree} implementation from the \texttt{scikit-learn} library.
For every couple $(\bar{x}, \bar{y})$ in the sample we computed the column density as
\begin{equation}
    \int_{-2r_\text{a}}^{2r_\text{a}} \rho(\bar{x}, \bar{y}, z) dz \approx \sum_{i=0}^{N_\text{z}-1} \rho(\bar{x}, \bar{y}, z_i) (z_{i+1}-z_i).
\end{equation}

\section{Resolution tests}
\label{app: res tests}
As a global convergence check, we show in Fig.\ref{fig:orbE_IE_relDiff} the energy budget in different resolutions. The orbital energy remains nearly identical across resolutions, which differ by less than $1\%$ for both the bound and unbound gas. The thermal energy exhibits differences that decrease over time: $\mathcal{R}$ peaks at early times, and later decreases around $1.1$. The radiation energy follows the same trend, with $\mathcal{R}\lesssim1.5$ at later times.

The increase of orbital kinetic energy for the unbound gas reflects in the efficiency of the outflow, whose mass loss rate increases as seen in Sec.\ref{sec: wind}. As this wind strengthens, it increasingly enclosed the photosphere, disregarding the resolution.
Figure \ref{fig:f_conv} shows the fraction $f$ of unbound lines of sight in different resolutions. Although material is more bound at higher resolution, the overall evolution indicates that the expanding photosphere progressively becomes unbound.

\bsp	
\label{lastpage}
\end{document}